\documentclass[journal]{IEEEtran}
\IEEEoverridecommandlockouts
\usepackage{cite}
\usepackage{booktabs}
\usepackage{amsmath,amssymb,amsfonts}
\usepackage{algorithmic}
\usepackage{graphicx}
\usepackage{subfigure}
\usepackage{textcomp}
\usepackage{tikz}
\usepackage{xcolor}
\usepackage{pgfplots}
\usepackage{pgfplotstable}
\usepackage{tikz,pgfplots}
\usetikzlibrary{positioning,shadows,shapes,backgrounds,calc}
\usepackage{pgfplotstable}
\pgfplotsset{compat=1.12} 

\usepackage{ragged2e}
\usepackage{booktabs,makecell,multirow,tabularx}
\usepackage{wrapfig}
\usepackage{dblfloatfix}
\usepackage{blindtext}
\usepackage{cuted}
\usepackage{subfigure}
\usetikzlibrary{spy}
\usepackage[pdftex]{hyperref} 
\hypersetup{colorlinks,linkcolor=red,citecolor=blue} 

\usepackage{balance}
\usepackage{pifont}
\usepackage{bbding}
\usepackage{fontawesome}

\newcounter{lemma}
\newtheorem{theorem}[lemma]{Theorem}

\newtheorem{lemma}{Lemma}

\newtheorem{exampleplain}{Example}

\newenvironment{proof}{{\indent  \it Proof:\quad}}{\hfill $\blacksquare$\par}

\definecolor{orange}{rgb}{1,0.7,0}
\definecolor{myDarkGreen}{rgb}{0.00000,0.58824,0.00000}%

\definecolor{myGreen}{rgb}{0.00000,0.49804,0.00000}

\newcommand{\changed}[1]{\textcolor{black}{#1}}
\newcommand{\ch}[1]{\textcolor{black}{#1}}

\begin{document}
\title{Analytical Model of Nonlinear Fiber Propagation for General Dual-Polarization Four-Dimensional Modulation Formats}

\author{Zhiwei Liang, Bin Chen,~\IEEEmembership{Senior Member,~IEEE}, Yi Lei,~\IEEEmembership{Member,~IEEE},\\ Gabriele Liga,~\IEEEmembership{Member,~IEEE} and Alex Alvarado,~\IEEEmembership{Senior Member,~IEEE}

\thanks{The work of B.~Chen, Y.~Lei and Z.~Liang are   supported by the National Natural Science Foundation of China  (No. 62171175 and 62001151), by the Fundamental Research Funds for the Central Universities (No. JZ2022HGTB0262, and by  Open Fund of State Key Laboratory of Information Photonics and Optical Communications (Beijing University of Posts and Telecommunications), P. R. China.
The work of G.~Liga is funded by the EuroTechPostdoc programme and  the European Research Council  under the European Union's Horizon 2020 research and innovation programme (No. 754462).
The work of A.~Alvarado is supported by the Netherlands Organisation for Scientific Research (NWO) via the VIDI Grant ICONIC (project number 15685). 
Parts of this paper have been presented at the \textit{European Conference Optical Communication (ECOC)}, Basel, Switzerland, 2022 \cite{zwLiang2022ECOC}. \textit{(Corresponding author: Bin Chen).}}
\thanks{Z.~Liang, B.~Chen, and Y.~Lei  are with School of Computer Science and Information Engineering, Hefei University of Technology,  Hefei, China and also with Intelligent Interconnected Systems Laboratory of Anhui Province (e-mails:~\{bin.chen,leiyi\}@hfut.edu.cn). 
}
\thanks{G.~Liga and A.~Alvarado are with  Department of Electrical Engineering, Eindhoven University of Technology, Eindhoven, The Netherlands (e-mails:~\{g.liga,a.alvarado\}@tue.nl).}
}




\maketitle
\begin{abstract}
Coherent dual-polarization (DP) optical transmission systems encode information on the four available degrees of freedom of an optical field: the two polarization states, each with two quadrature components. Such systems naturally operate based on a four-dimensional (4D) signal space. Having a general analytical model to accurately estimate nonlinear interference (NLI) is key to  analyze such transmission systems as well as to study how different DP-4D formats are affected by NLI. However, the available models in the literature are not completely general. They either do not apply to the entire DP-4D formats or do not consider all the NLI contributions. \ch{In this paper, we develop a model that applies to all DP-4D modulation formats with independent symbols.} Our model takes self-channel interference, cross-channel interference and multiple-channel interference effects into account. As an application of our model, we further study the effects of signal-noise interactions in long-haul transmission via the proposed model. When compared to previous results in the literature, our model is more accurate at predicting the contribution of NLI for both low and high dispersion fibers in single- and multi-channel transmission systems. For the NLI, we report an average gap from split step Fourier simulation results below 0.15~dB. \changed{The simulation results further show that by considering signal-noise interactions, the proposed model in long-haul transmission can reduce the transmission reach prediction error by 4\%.}
\end{abstract}

\begin{IEEEkeywords}
Nonlinear interference model, Four-dimensional modulation formats, Signal-noise interaction, Optical fiber communications
\end{IEEEkeywords}

\section{Introduction}
In optical communication systems, one of the main challenges is to make efficient use of existing network resources. 
To  achieve this, signal shaping has been investigated in optical fiber communications as an effective approach to achieve high spectral efficiencies (SEs). Shaping can be performed by changing the probability or position of the constellation points, which is known as probabilistic shaping (PS) \cite{Kschischang1993} and geometrical shaping (GS) \cite{ForneyJSAC1984}, resp.

\changed{Performing joint shaping over multiple dimensions, e.g.,  polarizations and  time slots \cite{Shiner:14,KoikeAkinoECOC2013}, wavelengths \cite{ErikssonECOC2013,Shen2022}, and cores \cite{ReneOFC2020}, to achieve large performance gains has received interest in the literature for both the additive white Gaussian noise (AWGN) \cite{Welti1974,Forney1989,AgrellJLT2009,Eric2022} and the optical fiber channel \cite{Kojima2017JLT,El-Rahman2017,Sebastiaan2022-4DGSS,Eric2022,Binchen2023JLT}.}  \changed{When constellation shaping tailored to the AWGN channel is used in the nonlinear optical channel, however, a nonlinear shaping gain penalty is introduced.} This penalty is caused by the modulation-dependent nonlinear interference (NLI). This effect was studied for example in \cite{Fehenberger16,Renner2017}.

In order to harvest most of the gains in the nonlinear fiber channel, heuristic ideas have been used in the literature. For example adding constant-modulus constraints\cite{Kojima2017JLT,BinChenJLT2019} or using shell shaping  \cite{Sebastiaan2022-4DGSS} in the optimization. Albeit such heuristics have the potential to significantly reduce the NLI, to fully maximize the performance in the nonlinear channel, an accurate analytical expression that links the modulation geometry and statistics to the induced NLI power is needed. As shown in \cite{Vinicius2021-4D1024} using the split-step Fourier method (SSFM) for constellation optimization becomes  computationally demanding  dimensions and moderate modulation cardinalities. Therefore, a general NLI model that allows fast and accurate estimation to capture the effect of  NLI is essential for optimizing and analyzing the performance of modulation formats in optical communication systems.

In the last two decades, many analytical nonlinear models have been presented in the literature. \changed{The models can be broadly grouped into time-domain and frequency-domain models \cite{Mecozzi2000,Louchet2003,Poggiolini2011,6276216,Carena:12,Mecozzi2012,Johannisson2013,Dar13,Carena14}. Some of the most prominent ones are Gaussian noise (GN) model and enhanced Gaussian noise (EGN) model, which are sufficiently accurate tool to predict the main system performance and  widely used in commercial fields. The GN model was derived based on the assumption that the signal statistically behaves as Gaussian noise over uncompensated links. However, soon after the introduction of the GN model, it was pointed out in \cite{Dar13,Carena14} that  ignoring all modulation-format-dependent terms leads to a substantial NLI overestimation up to several dB. }

To analyze and reduce the limitations of the GN model, a number of modulation format-dependent correction formulas have been proposed, effectively lifting the Gaussianity assumption of the transmitted signal. The first part of Table~\ref{tab:model_compare} shows the details of \emph{traditional} models for 2D modulation formats. As shown in Table~\ref{tab:model_compare}, the models in \cite{Dar13,Dar14} derived the correction terms including self-channel interference (SCI) and cross-phase modulation (XPM) in time domain. The model in \cite{Carena14} derived all the main NLI effects including SCI, cross-channel interference (XCI)\footnote{Recall that there are 4 XCI terms, often called X1, X2, X3, and X4 (see Fig.~\ref{fig:WDM} ahead, where X1 corresponds to XPM.}   and multiple-channel interference (MCI) were derived  in frequency domain. A major drawback of all these traditional models is that they can only be applied to polarization-multiplexed 2D (PM-2D) modulation formats, where two identical 2D formats are used to transmit information independently over the two orthogonal polarizations. PM-2D formats are only a subset of all possible dual-polarization four-dimensional (DP-4D) modulation formats. 

In order to fully explore the potential of DP-4D modulation formats in the nonlinear fiber channel, \cite{GabrieleEntropy2020} introduced the first 4D NLI model as a tool to efficiently trade-off linear and nonlinear shaping gains. The frequency-domain model in \cite{GabrieleEntropy2020} applies only to single-channel scenarios since it only considered SCI. Later in \cite{Hami2021}, a time-domain model  was introduced that considers both SCI and cross-phase modulation (XPM). The model in \cite{Hami2021} is  only valid for 4D symmetric constellations\footnote{Constellations which are symmetric with respect to the origin, and have the same power in both polarizations \cite[Sec.~I]{Hami2021}.} and  high-dispersion fiber systems (e.g., standard single mode fiber (SMF) in dispersion-uncompensated system). The model in \cite{Hami2021} was then extended to take SCI, XCI and MCI into account in \cite{Hami2022}.  The model in \cite{Hami2022} can be used for low dispersion fiber but still makes the assumption of 4D symmetric constellations. 
Recently, based on \cite{Hami2021}, a model for arbitrary 4D modulation formats was introduced in \cite{Hami2023JLT}. The model in \cite{Hami2023JLT} was derived in the time domain but only accounts partially for NLI effects (SCI and XPM terms only). The state-of-the-art NLI models for 4D modulation formats are summarized in  the second part of Table~\ref{tab:model_compare}.

The contribution of this paper is twofold. First, \ch{we derive an ``ultimate" 4D NLI model that covers all DP-4D  modulation formats with independent symbols.} We achieve this by extending the 4D NLI model (SCI-only) in \cite{GabrieleEntropy2020} to consider all the main NLI contributions, including SCI, XCI and MCI. Secondly, we extend our preliminary results in \cite{zwLiang2022ECOC} on signal-noise interactions for single-channel DP-4D systems to wavelength division multiplexed (WDM) systems. We perform a comprehensive numerical analysis in 
multi-channel WDM systems with three different fibers with different dispersion parameters. Our study is validated by analytically studying the effective signal-to-noise ratio (SNR) using general DP-4D formats. Our results show that the proposed 4D nonlinear model has a superior accuracy with a maximum deviation of 0.15~dB in terms of NLI power for all 4D modulation formats studied in this work. 
Moreover, by considering the signal-noise interactions in a multi-channel long-haul transmission system, the SNR estimation error can be reduced to 0.1~dB with respect to SSFM results, which can be translated into a 4\% prediction accuracy improvement in terms of transmission reach compared with only considering signal-signal interactions.

This paper is structured as follows. In Sec.~\ref{sec:system}, we  present the system model and review the effective SNR  considering the signal-signal and signal-noise interactions. Sec.~\ref{sec:model} presents the expression of the proposed NLI  model and the key steps of its derivation. Sec.~\ref{sec: results} is devoted to validate this model and assess the contribution of signal-noise interaction via a wide range of 4D modulation formats. Sec.~\ref{sec: conclusion} concludes this paper and outlines the potential direction for future research. Finally, the appendix provides a detailed derivation of the NLI model.

\begin{table}[!tb]
  \vspace{-1em}
\caption{NLI models for optical fiber communication systems relevant for this work.}
   \vspace{-0.5em}
\label{tab:model_compare}
    \centering
  {\renewcommand{\arraystretch}{1.03}
\centering
\begin{footnotesize}
\begin{tabular}{@{\hskip 0.2ex}c@{\hskip 0.5ex}|@{\hskip 0.8ex}c@{\hskip 0.8ex}|@{\hskip 0.5ex}c|@{\hskip 0.3ex}c|@{\hskip 0.3ex}c@{\hskip 0.5ex}|@{\hskip 0.8ex}c@{\hskip 0.8ex}|@{\hskip 0.8ex}c@{\hskip 0.2ex}}
\hline\hline
 {\textbf{Ref.}} & \textbf{SCI} & \textbf{XCI Terms} & \textbf{MCI} & \textbf{Mod. Format}  & \textbf{Dispersion} & \textbf{Dom.}\\
 \hline
\multicolumn{7}{c}{\textbf{2D NLI Model}}\\
 \hline
  \cite{Dar13,Dar14} & \checkmark  & XPM only & \ding{56}  & PM-2D & High  & $t$\\
   \hline
  \cite{Carena14}& \checkmark  & \checkmark(X1-X4) & \checkmark  & PM-2D & Any  & $f$ \\
\hline
\multicolumn{7}{c}{\textbf{4D NLI Model}}\\
\hline
\cite{GabrieleEntropy2020} & \checkmark & \ding{56} & \ding{56} & DP-4D & Any & $f$\\
\hline
\cite{Hami2021} & \checkmark & \checkmark {(XPM only)} & \ding{56} & Symmetric & High & $t$\\
\hline
\cite{Hami2022} & \checkmark & \checkmark(X1-X4) & \checkmark & Symmetric & Any & $t$\\
\hline
\cite{Hami2023JLT} & \checkmark & \checkmark {(XPM only)} & \ding{56} & DP-4D & High & $t$\\
\hline
\textbf{This work} & \checkmark & \checkmark(X1-X4) & \checkmark & DP-4D & Any & $f$\\
\hline\hline
\multicolumn{7}{l}{{$t$: time domain; $f$: frequency domain}; XPM same as X1 (see Fig.~\ref{fig:WDM})}

\end{tabular}
\end{footnotesize}

}
   \vspace{-1.8em}
\end{table}

 \begin{figure*}[!tb]
    \centering
      \centering
      \scalebox{0.95}{\includegraphics{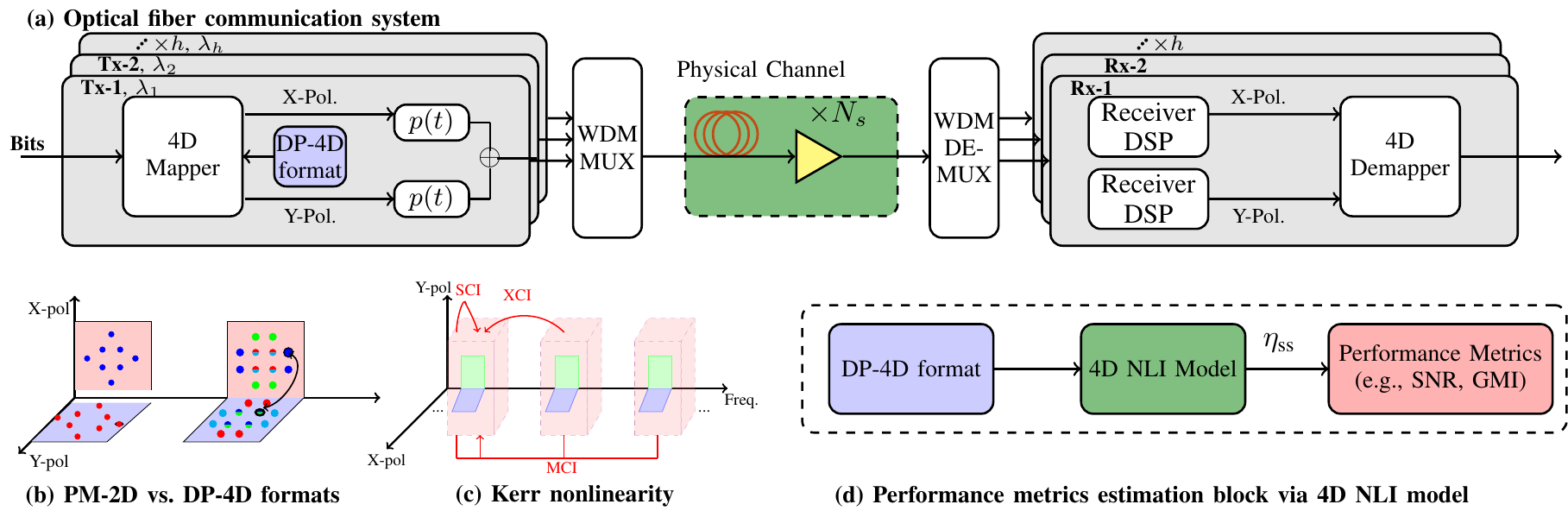}}
    \caption{(a) System model under investigation in this work which consists of a general WDM optical fiber system. (b) PM-2D vs. DP-4D formats. (c) An example of effects experienced by three frequency channels along  the modulated dimensions. (d) A general 
    block diagram using NLI model to estimate the modulation-dependent performance metrics.}
    \label{fig:System_Model}
    \vspace{-1em}
\end{figure*}

\section{System Model And Performance Metrics}\label{sec:system}
\subsection{System Model}
In this work, we consider the equivalent model of an optical fiber system shown in Fig.~\ref{fig:System_Model} (a). \changed{The physical channel is a multi-span, $h$-channel WDM fiber system using an ideal lumped amplification, for example Erbium-doped fiber amplifier (EDFA), per span to ideally compensate the span losses.} At the transmitter, the input bits are mapped into 4D symbols using a predefined DP-4D modulation format and its corresponding binary labeling. The 4D symbols are pulse-shaped by a real pulse $p(t)$ on x and y polarisation. The 4D transmitted WDM signal under a periodic assumption with period $T$ can be expressed in time domain as 
\changed{
\begin{align}\label{time-TxModel}
    E(t,0) = \sum_{h = -(N_{ch}-1)/2}^{(N_{ch}-1)/2}{\widetilde{E}(t,0)e^{jf_\text{c}^ht}},
\end{align}
where $N_{ch}$ (assumed odd) is the total number of channels and $f_\text{c}^h$ is the center frequency of h-th channel. According to the channel of interest (COI)  or  interference (INT) channel, the 4D transmitted signal over single channel $\widetilde{E}(t,0)$ can be represented as 
\vspace{-0.5em}
\begin{align}
    \widetilde{E}(t,0) = \sum_{k=-\infty}^{k=\infty}\boldsymbol{C}_{k}\delta(f-k\Delta_f),
\end{align}
in which
\begin{align}\label{Tx}
    & \boldsymbol{C}_k =\left\{
    \begin{aligned}
    & \Delta_fP(k\Delta_f)\sum_{n=-(W-1)/2}^{(W-1)/2}\boldsymbol{a}_n e^{-j2\pi\frac{kn}{W}}, h \text{ is COI}\\
    & \Delta_fP(k\Delta_f)\sum_{n=-(W-1)/2}^{(W-1)/2}\boldsymbol{b}_n e^{-j2\pi\frac{kn}{W}}, h \text{ is INT}
    \end{aligned},
    \right.
    \vspace{-0.5em}
\end{align}}
where  $T_s=1/R_s=T/W$ represents the symbol period. $R_s$ is the symbol rate and $W$ is the number of symbols transmitted every period $T$. The variable $\boldsymbol{a}_n$ or $\boldsymbol{b}_n$ denote that the sequence of symbols are from the COI or INT channel, respectively.

After transmitting the symbols over the physical channel, the received symbols are processed by a receiver DSP block, including chromatic dispersion compensation, matched filtering, sampling and  
ideal phase compensation for potential constant phase rotation. The symbols are then demapped by a 4D  demapper to generate soft information (i.e., log-likelihood ratios), which is then used to estimate the transmitted bits and to get the system performance metrics.

There are mainly two ways of generating a sequence of 4D symbols. These are schematically shown in the Fig.~\ref{fig:System_Model}~(b). The left hand side of Fig.~\ref{fig:System_Model}~(b) shows the case of PM-2D formats, where the 4D points can be described using independent and identically distributed 2D points on each polarization. The right hand side of Fig.~\ref{fig:System_Model}~(b) show the more general case called DP-4D, where the 2D constellations are jointly modulated on two orthogonal polarization state. In this case, the 2D projections in each polarization are not independent.

\subsection{Performance Metrics}
It is known that calculating performance metrics of optical transmission system using SSFM simulations is a time-consuming task. An NLI model can be an efficient way to solve this problem. The general idea is shown in Fig.~\ref{fig:System_Model} (d), where the NLI model is used to replace the time-consuming SSFM simulations in order to predict certain performance metrics. To explore the role of signal-noise interactions, in this paper we focus on predicting the effective SNR as well as the generalized mutual information (GMI) \cite{Alex2015,AIRAlex2018}. To estimate effective SNR, we will use our proposed model to estimate NLI power coefficients $\eta_{\text{ss}}$ and $\eta_{\text{sn}}$, which are associated to signal-signal (ss) and signal-noise (sn) components, respectively. As we will show later, these two coefficients are sufficient to estimate the effective SNR in a multi-channel multi-span scenario for arbitrary DP-4D modulation formats. To estimate the GMI, we will use our proposed model to predict the SNR for a given transmission scenario, and then use the Gauss-Hermite approximation to calculate the GMI.

Under an additive NLI noise assumption, the effective SNR over the two polarizations is defined as:
\begin{equation}\label{snr}
\begin{split}
    \text{SNR}^{\text{model}}_{\text{eff}} &\triangleq \frac{P}{N_s\sigma^2_{\text{ASE}}+\sigma^2_{\text{ss}}+\sigma^2_{\text{sn}}},
\end{split}
\end{equation}
where $P$ denotes the transmitted signal power per channel, $N_s$ is the number of spans. Assuming ideal transceivers, the main contributions to the total noise power are consisting of three parts: i) amplified spontaneous emission (ASE) noise over one span denoted as $\sigma^2_{\text{ASE}}$, ii) signal-signal NLI power denoted as $\sigma^2_{\text{ss}}$ and iii) signal-ASE noise NLI power denoted as $\sigma^2_{\text{sn}}$. The effective SNR in \eqref{snr} corresponds to the SNR after fiber propagation and the receiver DSP including chromatic dispersion compensation and phase compensation.

For dual-polarized signals over multi-span transmission, the signal-signal NLI power $\sigma^2_{\text{ss}}$ in \eqref{snr} can be approximated as \cite {7831073}
\begin{equation}\label{ss}
    \sigma^2_{\text{ss}}  \triangleq \sigma^2_{\text{ss,x}}+\sigma^2_{\text{ss,y}}\approx \tilde{\eta}_{\text{ss}}N_s^{1+\varepsilon}P^3 = \eta_{\text{ss}} P^3,
\end{equation}
where $\varepsilon$ is the so-called NLI coherence factor, which is a function of fiber link parameters (attenuation, dispersion, span length, etc) \cite[eq.~(40)]{Poggiolini2014} and
\begin{align}
\label{tilde.eta.ss}
\tilde{\eta}_{\text{ss}} & \triangleq \tilde{\eta}^{\text{x}}_{\text{ss}}+\tilde{\eta}^{\text{y}}_{\text{ss}}\\
\label{eta.ss}
\eta_{\text{ss}} & \triangleq \eta^{\text{x}}_{\text{ss}}+\eta^{\text{y}}_{\text{ss}}=\tilde{\eta}_{\text{ss}} N_s^{1+\varepsilon},
\end{align}
in which the $\tilde{\eta}_{\text{ss}}$ is the signal-signal NLI power coefficient (over one span). 
As shown in \eqref{ss}, from now on we will use  $\eta_{\text{ss}}$ to denote the accumulated signal-signal NLI power coefficient over  two polarizations and $N_s$ spans.

The ASE noise generated by the EDFAs leads not only to an additive white Gaussian noise (AWGN) but also to a nonlinear interference produced by the interaction of ASE noise and the transmitted signal \cite{7637002}. Under the assumption of a flat transmitted signal spectrum and same propagated signal and ASE noise bandwidth, the signal-ASE noise NLI power coefficient can be estimated as $\tilde{\eta}_{\text{sn}}=3\tilde{\eta}_{\text{ss}}$, where the $\tilde{\eta}_{\text{sn}}$ is the signal-ASE noise NLI power coefficient over one span  \cite[Sec.~3]{Cartledge:17}\cite[eq.~(1)]{7831073}. Thus, by following \cite [eq.~(8)]{Cartledge:17}, the NLI power of signal-ASE noise interactions  can be  expressed as 
\begin{align}\label{sn}    
\sigma^2_{\text{sn}}\triangleq \sigma^2_{\text{sn,x}}+\sigma^2_{\text{sn,y}}=\xi\tilde{\eta}_{\text{sn}}\sigma^2_{\text{ASE}}P^2=3\xi\tilde{\eta}_{\text{ss}}\sigma^2_{\text{ASE}}P^2,
\end{align}
where
\begin{align}\label{xi.exp}
    \xi = \sum_{n=1}^{N_s}n^{1+\varepsilon} \approx\frac{N_s^{2+\varepsilon}}{2+\varepsilon}+\frac{N_s^{1+\varepsilon}}{2}    
\end{align}
is the signal-ASE noise NLI accumulation coefficient \cite[Sec.~3]{Cartledge:17}, and $\tilde{\eta}_{\text{sn}}$ denotes the signal-ASE NLI power coefficient (over one span). 
Note that here the $\eta_{\text{sn}}$ is approximation, while the $\eta_{\text{ss}}$ do not use the same approximation but is computed exactly (in full integral form). 

The total NLI power is estimated using \eqref{ss} \eqref{eta.ss} \eqref{sn} and \eqref{xi.exp}. The effective SNR in \eqref{snr} can then be expressed as
\begin{align}\label{SNR}
    \begin{split}
    \text{SNR}^{\text{model}}_{\text{eff}} &\approx \frac{P}{N_s\sigma^2_{\text{ASE}}+\eta_{\text{ss}}P^3+3\eta_{\text{ss}}\bigl(\frac{N_s}{2+\varepsilon}+\frac{1}{2}\bigr)\sigma^2_{\text{ASE}}P^2}.
\end{split}
\end{align}

Note that $\eta_{\text{ss}}$ is a constant value (for a given system configuration) linked to the contributions of both modulation-independent and modulation-dependent nonlinearities. In  next section, the NLI power coefficient $\eta_{\text{ss}}$ will be shown including all the main NLI in multi-channel WDM optical systems. 
\begin{figure}[!tb]
\centering
\includegraphics{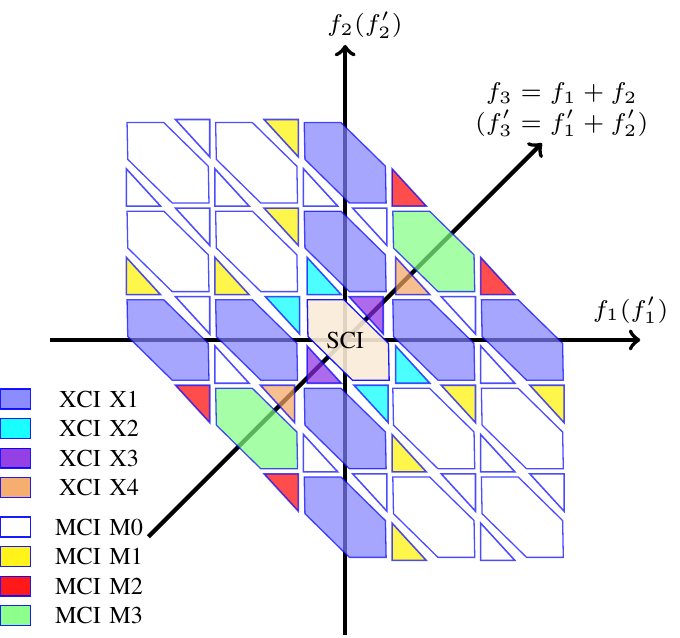}
    \caption{Example of regions of XCI and MCI for 5 channels at $f = 0$. 
    }
    \label{fig:WDM}
\vspace{-1em}
\end{figure}
\section{The Nonlinear model Derivation}\label{sec:model}
\changed{As shown in Fig.~\ref{fig:System_Model} (c), the NLI effects experienced by the transmitted signals in a multi-channel WDM optical system can be categorised using different names depending on the involved channels \cite{Carena2014}:}
\begin{itemize}
    \item SCI: NLI caused by the COI on itself.
    \item XCI: NLI affecting the COI caused by the beating of the COI with any single INT channel. XPM is a subset of the total XCI, namely the XCI contributions in X1.
    \item MCI: NLI affecting the COI, caused by the beating between frequency components located over two different INT channels or three  distinct channels.
\end{itemize}

\changed{The NLI effects can be physically interpreted as the frequency beating of the COI with all other channels through the four-wave mixing (FWM) process. This is reflected in the ``link function" which represents the normalized FWM efficiency $\mu(f_1,f_2,f)$ of the three ``pump" frequency beatings $f_1$, $f_2$ and $f_3=f_1+f_2-f$.} In the 4D model formulas, the power spectral density (PSD) of the NLI is provided by two triples: $(f_1,f_2,f_3)$ and $(f'_1,f'_2,f'_3)$. Since for each region in the $[f_1,f_2]$ plane there exist an equivalent region in the $[f'_1,f'_2]$ plane, integrating over  regions in the $[f_1,f_2]$ plane is enough \cite{Carena2014}. We will now provide a graphical intuitive description assuming $f = 0$ (for simplicity). In Fig.~\ref{fig:WDM}, we show the integration regions in the $[f_1,f_2]$ plane needed to obtain the power spectrum of NLI for $f = 0$, for a five-channel WDM system. Here, each island (including lozenge-shaped and triangles) represents a triple of frequencies, namely $(f_1,f_2,f_3)$. Based on the location of $(f_1,f_2,f_3)$ triples, all different NLI contributions can be fully categorized as XCI: X1, X2, X3 and X4 and MCI: M0, M1, M2 and M3. 

In this paper, we follow an approach similar to that of the EGN model \cite {Carena2014}, i.e., we derive the NLI power coefficient $\eta_{\text{ss}}$ in \eqref{eta.ss} by first expressing it as
\begin{align}
    \eta_{\text{ss}} =  \eta_{\text{ss,SCI}}+\eta_{\text{ss,XCI}}+\eta_{\text{ss,MCI}}.
\end{align}
The $\eta_{\text{ss,SCI}}$ term has been derived in \cite{GabrieleEntropy2020}. The other terms will be derived in the following subsections.

\subsection{SCI contribution}
In \cite{GabrieleEntropy2020}, a detailed derivation of the SCI term accounting for DP-4D  formats was shown, which has also been validated  in  \cite{9489833,zwLiang2022ECOC}. Here, we review the main defining formulas and key conclusions from \cite{GabrieleEntropy2020,9489833}. 

For general DP-4D formats, the modulation-dependent coefficient $\eta_{\text{ss,SCI}}=\eta_{\text{ss,SCI}}^{\text{x}}+\eta_{\text{ss,SCI}}^{\text{y}}$ for the x polarization can be calculated using \cite[eq.~(1)]{9489833}
\changed{\begin{align}\label{SCI}
\begin{split}
\eta_{\text{ss,SCI}}^{\text{x}} &\triangleq \frac{\sigma^2_{\text{SCI,x}}}{P^3} = P^{-3}\int_{-\infty}^{\infty}S^{\text{x}}_{\text{SCI}}(f,N_s,L_s)|P(f)|^2df,
\end{split}
\end{align}
where the $P(f)$ is the transmitted pulse spectrum and the SCI PSD $S^{\text{x}}_{\text{SCI}}(f,N_s,L_s)$ is given as
\begin{align}\label{SCI_PSD}
\begin{split}
S^{\text{x}}_{\text{SCI}}(f,N_s&,L_s) =  \left(\frac{8}{9}\right)^2\gamma^2\Bigg[R_s^3(\Phi_1\chi^{(1)}_{\text{SCI}}(f)+\Phi_2\chi^{(2)}_{\text{SCI}}(f)\\
&+\Phi_3\chi^{(3)}_{\text{SCI}}(f)+R_s^2(\Psi_1\chi^{(4)}_{\text{SCI}}(f)
    +2\Re\{\Psi_2\chi^{(5)}_{\text{SCI}}(f)\\
    &+\Psi_3\chi^{(5)}_{\text{SCI}}(f)^*\}+\Psi_4\chi^{(6)}_{\text{SCI}}(f)+2\Re\{\Lambda_1\chi^{(7)}_{\text{SCI}}(f)\\
    &+\Lambda_2\chi^{(7)}_{\text{SCI}}(f)^*\}+\Lambda_3\chi^{(8)}_{\text{SCI}}(f)+2\Re\{\Lambda_4\chi^{(9)}_{\text{SCI}}(f)\\
    &+\Lambda_5\chi^{(9)}_{\text{SCI}}(f)^*\}+\Lambda_6\chi^{(10)}_{\text{SCI}}(f))+R_s\Xi_1\chi^{(11)}_{\text{SCI}}(f)\Bigg],
    \end{split}
\end{align}}
where $\gamma$ is the nonlinear coefficient, $R_s$ is the symbol rate, $\Re\{\cdot\}$ denote the real part of a complex number,  the coefficients $\Phi_k, k=1,2,3, \Psi_k, k=1,2,3,4, \Lambda_k, k=1,2,...,6$, and $\Xi_1$ are modulation format-dependent terms as  functions of several different intra- and cross- polarization moments, which can be found in Table~\ref{tab:two}. The coefficients \changed{$\chi^{(k)}_{\text{SCI}}, k=1,2,...,11$} are given in Table~\ref{tab:three} and are the frequency-dependent integrals over the channel bandwidth. \changed{The outer boundaries of the integration domain is explained in Appendix C.} These coefficients are independent with the shape of the  modulation format. The expression for $\eta_{\text{ss,SCI}}^{\text{y}}$ can be obtained applying the transformation $\text{x} \rightarrow \text{y}$ and $\text{y} \rightarrow \text{x}$ to \eqref{SCI}.

\subsection{XCI contribution}
\changed{The XCI contributions of multi-channel WDM systems can be added up independently, and therefore, the total XCI is then the sum of the contributions of each INT channel that exists in the WDM channels. The total XCI coefficient $\eta_{\text{ss,XCI}}^{\text{x}}$ for the x polarization can be shown as
\begin{align}\label{XCI_tot}
\eta_{\text{ss,XCI}}^{\text{x}} =  \sum_{\substack{h = -(N_{ch}-1)/2 \\ h \neq 0}}^{(N_{ch}-1)/2}   \eta_{\text{ss,XCI}}^{\text{x},h},
\end{align}
}

\changed{In addition, to generalise this result to aperiodic signal, we follows the same approach in \cite{GabrieleEntropy2020}, i.e., by  setting the period $T$ go to infinity. The following Theorem presents the XCI contributions.}

\begin{table*}[!htbp]
        \centering
		\caption{Modulation format-dependent coefficients in \eqref{SCI_PSD}, \eqref{XCI_PSD} and \eqref{MCI_PSD}}
  \renewcommand\arraystretch{2}
		\scalebox{0.95}{
		\begin{tabular}{c|c|c|c|c}
\hline
\hline
 Name & Value & SCI & XCI & MCI \\
\hline
\hline
$\Phi_{1}$ & 
\makecell[l]{$2 \mathbb{E}^{3}\{\left|a_{\text{x}}\right|^{2}\}+4 \mathbb{E}\{\left|a_{\text{x}}\right|^{2}\}\left|\mathbb{E}\{a_{\text{x}} a_{\text{y}}^{*}\right\}|^{2}+\mathbb{E}\{\left|a_{\text{x}}\right|^{2}\} \mathbb{E}^{2}\{\left|a_{\text{y}}\right|^{2}\}+\left|\mathbb{E}\{a_{\text{x}} a_{\text{y}}^{*}\right\}|^{2} \mathbb{E}\{\left|a_{\text{y}}\right|^{2}\}$ 
} & \checkmark &  & \\
\hline
 $\Phi_{2}$ & 
\makecell[l]{$4 \mathbb{E}\{\left|a_{\text{x}}\right|^{2}\}\left|\mathbb{E}\{a_{\text{x}}^{2}\}\right|^{2}+\mathbb{E}\{\left|a_{\text{x}}\right|^{2}\}\left|\mathbb{E}\{a_{\text{y}}^{2}\}\right|^{2}+4 \mathbb{E}\{\left|a_{\text{x}}\right|^{2}\}\left|\mathbb{E}\{a_{\text{x}} a_{\text{y}}\}\right|^{2}+\left|\mathbb{E}\{a_{\text{x}} a_{\text{y}}\}\right|^{2} \mathbb{E}\{\left|a_{\text{y}}\right|^{2}\}$ \\
$ +2 \Re\{\mathbb{E}\left\{a_{\text{x}} a_{\text{y}}\} \mathbb{E}\{a_{\text{x}}^{*} a_{\text{y}}\} \mathbb{E}^{*}\{a_{\text{y}}^{2}\}+2 \mathbb{E}^{*}\{a_{\text{x}}^{2}\} \mathbb{E}\{a_{\text{x}} a_{\text{y}}\} \mathbb{E}\{a_{\text{x}} a_{\text{y}}^{*}\right\}\}$
} & \checkmark &  & \\
\hline
$\Phi_{3}$ & 
\makecell[l]{$\mathbb{E}\{\left|a_{\text{x}}\right|^{2}\}\left|\mathbb{E}\{a_{\text{x}}^{2}\}\right|^{2}+\left|\mathbb{E}\{a_{\text{x}} a_{\text{y}}\}\right|^{2} \mathbb{E}\{\left|a_{\text{y}}\right|^{2}\}+2 \Re\{\mathbb{E}\{a_{\text{x}}^{2}\} \mathbb{E}^{*}\{a_{\text{x}} a_{\text{y}}\} \mathbb{E}\{a_{\text{x}}^{*} a_{\text{y}}\}\}$ 
}& \checkmark &  & \\
\hline

$\Psi_{1}$ & 
\makecell[l]{$4|\mathbb{E}\{a_{\text{x}}\left|a_{\text{x}}\right|^{2}\}|^{2}+4|\mathbb{E}\{\left|a_{\text{x}}\right|^{2} a_{\text{y}}\}|^{2}+\mathbb{E}\{\left|a_{\text{x}}\right|^{2} a_{\text{y}}\} \mathbb{E}\{a_{\text{y}}^{*}\left|a_{\text{y}}\right|^{2}\}+\mathbb{E}\{\left|a_{\text{x}}\right|^{2} a_{\text{y}}^{*}\} \mathbb{E}\{a_{\text{y}}\left|a_{\text{y}}\right|^{2}\}+|\mathbb{E}\{a_{\text{x}}\left|a_{\text{y}}\right|^{2}\}|^{2}$\\
$ +|\mathbb{E}\{a_{\text{x}}^{*} a_{\text{y}}^{2}\}|^{2}+2 \Re\{\mathbb{E}\{a_{\text{x}}^{*}\left|a_{\text{x}}\right|^{2}\} \mathbb{E}\{a_{\text{x}}\left|a_{\text{y}}\right|^{2}\}\}$ 
}& \checkmark &  & \\
\hline
$\Psi_{2}$ & 
\makecell[l]{$2|\mathbb{E}\{a_{\text{x}}\left|a_{\text{x}}\right|^{2}\}|^{2}+2|\mathbb{E}\{\left|a_{\text{x}}\right|^{2} a_{\text{y}}\}|^{2}+\mathbb{E}\{\left|a_{\text{x}}\right|^{2} a_{\text{y}}^{*}\} \mathbb{E}\{a_{\text{y}}\left|a_{\text{y}}\right|^{2}\}+|\mathbb{E}\{a_{\text{x}}\left|a_{\text{y}}\right|^{2}\}|^{2}$ 
}& \checkmark &  & \\
\hline
$\Psi_{3}$ & 
\makecell[l]{$\mathbb{E}\{a_{\text{x}}^{*}\left|a_{\text{x}}\right|^{2}\} \mathbb{E}\{a_{\text{x}}\left|a_{\text{y}}\right|^{2}\}+\left|\mathbb{E}\{a_{\text{x}}^{2} a_{\text{y}}^{*}\}\right|^{2}$ 
}& \checkmark &  & \\
\hline
$\Psi_{4}$ & 
\makecell[l]{$\left|\mathbb{E}\{a_{\text{x}}^{3}\}\right|^{2}+2\left|\mathbb{E}\{a_{\text{x}}^{2} a_{\text{y}}\}\right|^{2}+\left|\mathbb{E}\{a_{\text{x}} a_{\text{y}}^{2}\}\right|^{2}$ 
}& \checkmark &  & \\
\hline

$\Lambda_{1}$ &
\makecell[l]{$-3 \mathbb{E}\{\left|a_{\text{x}}\right|^{2}\}\left|\mathbb{E}\{a_{\text{x}}^{2}\}\right|^{2}+\mathbb{E}^{*}\{a_{\text{x}}^{2}\left|a_{\text{x}}\right|^{2}\} \mathbb{E}\{a_{\text{x}}^{2}\}-\left|\mathbb{E}\{a_{\text{x}}^{2}\}\right|^{2} \mathbb{E}\{\left|a_{\text{y}}\right|^{2}\}-2\left|\mathbb{E}\{a_{\text{x}} a_{\text{y}}\}\right|^{2} \mathbb{E}\{\left|a_{\text{y}}\right|^{2}\}$ \\ $+\mathbb{E}\{a_{\text{x}}^{2}\} \mathbb{E}^{*}\{a_{\text{x}}^{2}\left|a_{\text{y}}\right|^{2}\}-2 \mathbb{E}\{a_{\text{x}}^{2}\} \mathbb{E}^{*}\{a_{\text{x}} a_{\text{y}}\} \mathbb{E}\{a_{\text{x}}^{*} a_{\text{y}}\}+\mathbb{E}\{a_{\text{x}} a_{\text{y}}\} \mathbb{E}^{*}\{a_{\text{x}} a_{\text{y}}\left|a_{\text{y}}\right|^{2}\}-\mathbb{E}\{a_{\text{x}} a_{\text{y}}\} \mathbb{E}\{a_{\text{x}}^{*} a_{\text{y}}\} \mathbb{E}^{*}\{a_{\text{y}}^{2}\}$ 
}& \checkmark &  & \\
\hline
$\Lambda_{2}$ &
\makecell[l]{$-2 \mathbb{E}\{\left|a_{\text{x}}\right|^{2}\}\left|\mathbb{E}\{a_{\text{x}} a_{\text{y}}\}\right|^{2}+\mathbb{E}\{a_{\text{x}} a_{\text{y}}\} \mathbb{E}^{*}\{a_{\text{x}} a_{\text{y}}\left|a_{\text{x}}\right|^{2}\}-\mathbb{E}\{a_{\text{x}}^{2}\} \mathbb{E}^{*}\{a_{\text{x}} a_{\text{y}}\} \mathbb{E}\{a_{\text{x}}^{*} a_{\text{y}}\}$ 
}& \checkmark &  & \\
\hline
$\Lambda_{3}$ &
\makecell[l]{$4 \mathbb{E}\{\left|a_{\text{x}}\right|^{4}\} \mathbb{E}\{\left|a_{\text{x}}\right|^{2}\}-4 \mathbb{E}\{\left|a_{\text{x}}\right|^{2}\}\left|\mathbb{E}\{a_{\text{x}}^{2}\}\right|^{2}-8 \mathbb{E}^{3}\{\left|a_{\text{x}}\right|^{2}\}+4 \mathbb{E}\{\left|a_{\text{x}}\right| 2^{2}\} \mathbb{E}\{\left|a_{\text{x}}\right|^{2}\left|a_{\text{y}}\right|^{2}\}-12 \mathbb{E}\{\left|a_{\text{x}}\right|^{2}\}\left|\mathbb{E}\{a_{\text{x}} a_{\text{y}}^{*}\}\right|^{2}$ \\
$ -4 \mathbb{E}\{\left|a_{\text{x}}\right|^{2}\}\left|\mathbb{E}\{a_{\text{x}} a_{\text{y}}\}\right|^{2}-4 \mathbb{E}^{2}\{\left|a_{\text{x}}\right|^{2}\} \mathbb{E}\{\left|a_{\text{y}}\right|^{2}\}-3 \mathbb{E}\{\left|a_{\text{x}}\right|^{2}\} \mathbb{E}^{2}\{\left|a_{\text{y}}\right|^{2}\}-\mathbb{E}\{\left|a_{\text{x}}\right|^{2}\}\left|\mathbb{E}\{a_{\text{y}}^{2}\}\right|^{2}+\mathbb{E}\{\left|a_{\text{x}}\right|^{2}\left|a_{\text{y}}\right|^{2}\} \mathbb{E}\{\left|a_{\text{y}}\right|^{2}\} $\\
$+\mathbb{E}\{\left|a_{\text{x}}\right|^{2}\} \mathbb{E}\{\left|a_{\text{y}}\right|^{4}\}-5\left|\mathbb{E}\{a_{\text{x}} a_{\text{y}}^{*}\}\right|^{2} \mathbb{E}\{\left|a_{\text{y}}\right|^{2}\}-\left|\mathbb{E}\{a_{\text{x}} a_{\text{y}}\}\right|^{2} \mathbb{E}\{\left|a_{\text{y}}\right|^{2}\}+2 \Re\left\{2 \mathbb{E}\{a_{\text{x}} a_{\text{y}}^{*}\} \mathbb{E}\{a_{\text{x}}^{*} a_{\text{y}}\left|a_{\text{x}}\right|^{2}\}\right.$ \\ $\left.-\mathbb{E}\{a_{\text{x}} a_{\text{y}}\} \mathbb{E}\{a_{\text{x}}^{*} a_{\text{y}}\}\mathbb{E}^{*}\{a_{\text{y}}^{2}\}+\mathbb{E}\{a_{\text{x}}^{*} a_{\text{y}}\} \mathbb{E}\{a_{\text{x}} a_{\text{y}}^{*}\left|a_{\text{y}}\right|^{2}\}-2 \mathbb{E}^{*}\{a_{\text{x}}^{2}\} \mathbb{E}\{a_{\text{x}} a_{\text{y}}\} \mathbb{E}\{a_{\text{x}} a_{\text{y}}^{*}\}\right\}$ 
}& \checkmark &  & \\
\hline
$\Lambda_{4}$ & 
\makecell[l]{$-6 \mathbb{E}\{\left|a_{\text{x}}\right|^{2}\}\left|\mathbb{E}\{a_{\text{x}}^{2}\}\right|^{2}+2 \mathbb{E}^{*}\{a_{\text{x}}^{2}\left|a_{\text{x}}\right|^{2}\} \mathbb{E}\{a_{\text{x}}^{2}\}+4 \mathbb{E}\{\left|a_{\text{x}}\right|^{2}\}\left|\mathbb{E}\{a_{\text{x}} a_{\text{y}}\}\right|^{2}-\mathbb{E}\{\left|a_{\text{x}}\right|^{2}\}\left|\mathbb{E}\{a_{\text{y}}^{2}\}\right|^{2}$ \\
$ +\mathbb{E}^{*}\{\left|a_{\text{x}}\right|^{2} a_{\text{y}}^{2}\} \mathbb{E}\{a_{\text{y}}^{2}\}+2 \mathbb{E}\{a_{\text{x}} a_{\text{y}}\} \mathbb{E}^{*}\{a_{\text{x}}\left|a_{\text{x}}\right|^{2} a_{\text{y}}\}-2\left|\mathbb{E}\{a_{\text{x}} a_{\text{y}}\}\right|^{2} \mathbb{E}\{\left|a_{\text{y}}\right|^{2}\}-2 \mathbb{E}^{*}\{a_{\text{x}}^{2}\} \mathbb{E}\{a_{\text{x}} a_{\text{y}}\} \mathbb{E}\{a_{\text{x}} a_{\text{y}}^{*}\} $\\ 
$+\mathbb{E}\{a_{\text{x}} a_{\text{y}}\} \mathbb{E}^{*}\{a_{\text{x}} a_{\text{y}}\left|a_{\text{y}}\right|^{2}\}-\mathbb{E}^{*}\{a_{\text{x}} a_{\text{y}}\} \mathbb{E}\{a_{\text{x}} a_{\text{y}}^{*}\} \mathbb{E}\{a_{\text{y}}^{2}\}-2 \Re\{\mathbb{E}^{*}\{a_{\text{x}} a_{\text{y}}\} \mathbb{E}\{a_{\text{x}} a_{\text{y}}^{*}\} \mathbb{E}\{a_{\text{y}}^{2}\}\}$ 
}& \checkmark &  & \\
\hline
$\Lambda_{5}$ & 
\makecell[l]{$-2 \mathbb{E}\{\left|a_{\text{x}}\right|^{2}\}\left|\mathbb{E}\{a_{\text{x}} a_{\text{y}}\}\right|^{2}+\mathbb{E}\{a_{\text{x}} a_{\text{y}}\} \mathbb{E}^{*}\{a_{\text{x}} a_{\text{y}}\left|a_{\text{x}}\right|^{2}\}-\left|\mathbb{E}\{a_{\text{x}}^{2}\}\right|^{2} \mathbb{E}\{\left|a_{\text{y}}\right|^{2}\}-\mathbb{E}^{*}\{a_{\text{x}}^{2}\} \mathbb{E}\{a_{\text{x}} a_{\text{y}}\} \mathbb{E}\{a_{\text{x}} a_{\text{y}}^{*}\}$ \\
$ -2 \Re\{\mathbb{E}\{a_{\text{x}}^{2}\} \mathbb{E}^{*}\{a_{\text{x}} a_{\text{y}}\} \mathbb{E}\{a_{\text{x}}^{*} a_{\text{y}}\}\}$ 
}& \checkmark &  & \\
\hline
$\Lambda_{6}$ & 
\makecell[l]{$-2 \mathbb{E}^{3}\{\left|a_{\text{x}}\right|^{2}\}+\mathbb{E}\{\left|a_{\text{x}}\right|^{4}\} \mathbb{E}\{\left|a_{\text{x}}\right|^{2}\}-\mathbb{E}\{\left|a_{\text{x}}\right|^{2}\}\left|\mathbb{E}\{a_{\text{x}}^{2}\}\right|^{2}-4 \mathbb{E}\{\left|a_{\text{x}}\right|^{2}\}\left|\mathbb{E}\{a_{\text{x}} a_{\text{y}}^{*}\}\right|^{2}-\mathbb{E}\{\left|a_{\text{x}}\right|^{2}\} \mathbb{E}^{2}\{\left|a_{\text{y}}\right|^{2}\}$ \\ $+\mathbb{E}\{\left|a_{\text{x}}\right|^{2}\left|a_{\text{y}}\right|^{2}\} \mathbb{E}\{\left|a_{\text{y}}\right|^{2}\}-\left|\mathbb{E}\{a_{\text{x}} a_{\text{y}}^{*}\}\right|^{2} \mathbb{E}\{\left|a_{\text{y}}\right|^{2}\}-\left|\mathbb{E}\{a_{\text{x}} a_{\text{y}}\}\right|^{2} \mathbb{E}\{\left|a_{\text{y}}\right|^{2}\} $\\
$ +2 \Re\{\mathbb{E}\{a_{\text{x}} a_{\text{y}}^{*}\} \mathbb{E}\{a_{\text{x}}^{*} a_{\text{y}}\left|a_{\text{x}}\right|^{2}\}-\mathbb{E}\{a_{\text{x}}^{2}\} \mathbb{E}^{*}\{a_{\text{x}} a_{\text{y}}\} \mathbb{E}\{a_{\text{x}}^{*} a_{\text{y}}\}\}$ 
}& \checkmark &  & \\
\hline

$\Xi_{1}$ &
\makecell[l]{$\mathbb{E}\{\left|a_{\text{x}}\right|^{6}\}-9 \mathbb{E}\{\left|a_{\text{x}}\right|^{4}\} \mathbb{E}\{\left|a_{\text{x}}\right|^{2}\}+12 \mathbb{E}^{3}\{\left|a_{\text{x}}\right|^{2}\}-2 \mathbb{E}\{\left|a_{\text{x}}\right|^{4}\} \mathbb{E}\{\left|a_{\text{y}}\right|^{2}\}+\mathbb{E}\{\left|a_{\text{x}}\right|^{2}\left|a_{\text{y}}\right|^{4}\}-8 \mathbb{E}\{\left|a_{\text{x}}\right|^{2}\} \mathbb{E}\{\left|a_{\text{x}}\right|^{2}\left|a_{\text{y}}\right|^{2}\}$ \\
$ -4 \mathbb{E}\{\left|a_{\text{x}}\right|^{2}\left|a_{\text{y}}\right|^{2}\} \mathbb{E}\{\left|a_{\text{y}}\right|^{2}\}+2 \mathbb{E}\{\left|a_{\text{x}}\right|^{4}\left|a_{\text{y}}\right|^{2}\}-\mathbb{E}\{\left|a_{\text{x}}\right|^{2}\} \mathbb{E}\{\left|a_{\text{y}}\right|^{4}\}+4 \mathbb{E}\{\left|a_{\text{x}}\right|^{2}\} \mathbb{E}^{2}\{\left|a_{\text{y}}\right|^{2}\}+8 \mathbb{E}^{2}\{\left|a_{\text{x}}\right|^{2}\} \mathbb{E}\{\left|a_{\text{y}}\right|^{2}\}$ \\
$ +18 \mathbb{E}\{\left|a_{\text{x}}\right|^{2}\}\left|\mathbb{E}\{a_{\text{x}}^{2}\}\right|^{2}-\left|\mathbb{E}\{a_{\text{x}}^{3}\}\right|^{2}-9|\mathbb{E}\{a_{\text{x}}\left|a_{\text{x}}\right|^{2}\}|^{2}+2 \mathbb{E}\{\left|a_{\text{x}}\right|^{2}\}\left|\mathbb{E}\{a_{\text{y}}^{2}\}\right|^{2}-4|\mathbb{E}\{a_{\text{x}}\left|a_{\text{y}}\right|^{2}\}|^{2}-8|\mathbb{E}\{\left|a_{\text{x}}\right|^{2} a_{\text{y}}\}|^{2}$ \\ 
$+8\left|\mathbb{E}\{a_{\text{x}} a_{\text{y}}^{*}\}\right|^{2} \mathbb{E}\{\left|a_{\text{y}}\right|^{2}\}+8\left|\mathbb{E}\{a_{\text{x}} a_{\text{y}}\}\right|^{2} \mathbb{E}\{\left|a_{\text{y}}\right|^{2}\}-\left|\mathbb{E}\{a_{\text{x}} a_{\text{y}}^{2}\}\right|^{2}-\left|\mathbb{E}\{a_{\text{x}}^{*} a_{\text{y}}^{2}\}\right|^{2}+16 \mathbb{E}\{\left|a_{\text{x}}\right|^{2}\}\left|\mathbb{E}\{a_{\text{x}} a_{\text{y}}^{*}\}\right|^{2}$ 
\\
$ -2\left|\mathbb{E}\{a_{\text{x}}^{2} a_{\text{y}}^{*}\}\right|^{2}+16 \mathbb{E}\{\left|a_{\text{x}}\right|^{2}\}\left|\mathbb{E}\{a_{\text{x}} a_{\text{y}}\}\right|^{2}+4\left|\mathbb{E}\{a_{\text{x}}^{2}\}\right|^{2} \mathbb{E}\{\left|a_{\text{y}}\right|^{2}\}-2\left|\mathbb{E}\{a_{\text{x}}^{2} a_{\text{y}}\}\right|^{2}+2 \Re\left\{4 \mathbb{E}\{a_{\text{x}} a_{\text{y}}\} \mathbb{E}\{a_{\text{x}}^{*} a_{\text{y}}\} \mathbb{E}^{*}\{a_{\text{y}}^{2}\}\right. $
\\
$ -3 \mathbb{E}\{a_{\text{x}}^{2}\left|a_{\text{x}}\right|^{2}\} \mathbb{E}^{*}\{a_{\text{x}}^{2}\}-2 \mathbb{E}\{\left|a_{\text{x}}\right|^{2} a_{\text{y}}\} \mathbb{E}\{a_{\text{y}}^{*}\left|a_{\text{y}}\right|^{2}\}-\mathbb{E}\{\left|a_{\text{x}}\right|^{2} a_{\text{y}}^{2}\} \mathbb{E}^{*}\{a_{\text{y}}^{2}\}-2 \mathbb{E}\{a_{\text{x}} a_{\text{y}}\} \mathbb{E}^{*}\{a_{\text{x}} a_{\text{y}}\left|a_{\text{y}}\right|^{2}\} $
\\$ -\mathbb{E}\{a_{\text{x}} a_{\text{y}}^{*}\} \mathbb{E}\{a_{\text{x}}^{*} a_{\text{y}}\left|a_{\text{y}}\right|^{2}\}-2 \mathbb{E}\{a_{\text{x}}^{2}\} \mathbb{E}^{*}\{a_{\text{x}}^{2}\left|a_{\text{y}}\right|^{2}\}-\mathbb{E}\{a_{\text{x}}\left|a_{\text{x}}\right|^{2}\} \mathbb{E}\{a_{\text{x}}\left|a_{\text{y}}\right|^{2}\}-4 \mathbb{E}\{a_{\text{x}} a_{\text{y}}^{*}\} \mathbb{E}\{a_{\text{x}}^{*} a_{a_{\text{y}}} a_{\text{x}} 2^{2}\} $\\
$ \left.-4 \mathbb{E}\{a_{\text{x}} a_{\text{y}}\} \mathbb{E}^{*}\{a_{\text{x}} a_{\text{y}}\left|a_{\text{x}}\right|^{2}\}+8 \mathbb{E}\{a_{\text{x}}^{2}\} \mathbb{E}^{*}\{a_{\text{x}} a_{\text{y}}\} \mathbb{E}\{a_{\text{x}}^{*} a_{\text{y}}\}\right\}$ 
}& \checkmark &  & \\
\hline
$\Phi_4$ & 
\makecell[l]{$4\mathbb{E}\{|a_\text{x}|^2 \}\mathbb{E}^2\{|b_\text{x}|^2 \}+\mathbb{E}\{|a_\text{y}|^2 \}\mathbb{E}\{|b_\text{x}|^2 \}\mathbb{E}\{|b_\text{y}|^2 \}+4\mathbb{E}\{|a_\text{x}|^2 \}|\mathbb{E}\{b_\text{x}^{\ast}b_\text{y} \}|^2+\mathbb{E}\{|a_\text{x}|^2 \}\mathbb{E}^2\{|b_\text{y}|^2 \}$\\
$+2\Re\{2\mathbb{E}\{a_\text{x}a_\text{y}^{\ast} \}\mathbb{E}\{|b_\text{x}|^2 \}\mathbb{E}\{b_\text{x}^{\ast}b_\text{y} \}+\mathbb{E}\{a_\text{x}a_\text{y}^{\ast} \}\mathbb{E}\{|b_\text{y}|^2 \}\mathbb{E}\{b_\text{x}^{\ast}b_\text{y}\} \}$
}&  & \checkmark & \checkmark\\
\hline
$\Phi_5$ & 
\makecell[l]{
$4\mathbb{E}\{|a_\text{x}|^2 \}|\mathbb{E}\{b_\text{x}b_\text{y} \}|^2+\mathbb{E}\{|a_\text{y}|^2 \}|\mathbb{E}\{b_\text{x}b_\text{y} \}|^2+4\mathbb{E}\{|a_\text{x}|^2 \}|\mathbb{E}\{b_\text{x}^2 \}|^2+\mathbb{E}\{|a_\text{x}|^2 \}|\mathbb{E}\{b_\text{y}^2 \}|^2$\\
$+2\Re\{2\mathbb{E}\{a_\text{x}a_\text{y}^{\ast} \}\mathbb{E}^{\ast}\{b_\text{x}^2 \}\mathbb{E}\{b_\text{x}b_\text{y} \}+\mathbb{E}\{a_\text{x}a_\text{y}^{\ast} \}\mathbb{E}\{b_\text{y}^2 \}\mathbb{E}^{\ast}\{b_\text{x}b_\text{y}\} \}$
}&  & \checkmark & \checkmark\\
\hline
$\Phi_6 $ &  			
\makecell[l]{
$4\mathbb{E}\{|a_\text{x}|^2 \}\mathbb{E}\{|b_\text{x}|^4 \}-8\mathbb{E}\{|a_\text{x}|^2 \}\mathbb{E}^2\{|b_\text{x}|^2 \}-4\mathbb{E}\{|a_\text{x}|^2 \}|\mathbb{E}\{b_\text{x}^2 \}|^2-\mathbb{E}\{|a_\text{y}|^2 \}|\mathbb{E}\{b_\text{x}b^{\ast}_\text{y} \}|^2-\mathbb{E}\{|a_\text{y}|^2 \}|\mathbb{E}\{b_\text{x}b_\text{y} \}|^2$\\
$+\mathbb{E}\{|a_\text{x}|^2 \}\mathbb{E}\{|b_\text{y}|^4 \}-2\mathbb{E}\{|a_\text{x}|^2 \}\mathbb{E}^2\{|b_\text{y}|^2 \}-\mathbb{E}\{|a_\text{x}|^2 \}|\mathbb{E}\{b_\text{y}^2 \}|^2-2\mathbb{E}\{|a_\text{x}|^2 \}|\mathbb{E}\{b^{\ast}_\text{x}b_\text{y} \}|^2-2\mathbb{E}\{|a_\text{x}|^2 \}|\mathbb{E}\{b_\text{x}b_\text{y}^{\ast} \}|^2$\\
$-4\mathbb{E}\{|a_\text{x}|^2 \}|\mathbb{E}\{b_\text{x}b_\text{y} \}|^2+4\mathbb{E}\{|a_\text{x}|^2 \}\mathbb{E}\{|b_\text{x}|^2|b_\text{y}|^2 \}+\mathbb{E}\{|a_\text{y}|^2 \}\mathbb{E}\{|b_\text{x}|^2|b_\text{y}|^2 \}-4\mathbb{E}\{|a_\text{x}|^2 \}\mathbb{E}\{|b_\text{x}|^2\}\mathbb{E}\{|b_\text{y}|^2 \}$\\
$-\mathbb{E}\{|a_\text{y}|^2 \}\mathbb{E}\{|b_\text{x}|^2\}\mathbb{E}\{|b_\text{y}|^2 \}+2\Re\{-2\mathbb{E}\{a_\text{x}a^{\ast}_\text{y} \}\mathbb{E}\{b_\text{x}b_\text{y} \}\mathbb{E}^{\ast}\{b_\text{x}^2 \}-\mathbb{E}\{a_\text{x}a^{\ast}_\text{y} \}\mathbb{E}\{b_\text{y}^2 \}\mathbb{E}^{\ast}\{b_\text{x}b_\text{y} \}$\\
$-2\mathbb{E}\{a_\text{x}a^{\ast}_\text{y} \}\mathbb{E}\{|b_\text{x}|^2 \}\mathbb{E}\{b_\text{x}^{\ast}b_\text{y} \}-\mathbb{E}\{a_\text{x}a^{\ast}_\text{y} \}\mathbb{E}\{|b_\text{y}|^2 \}\mathbb{E}\{b_\text{x}^{\ast}b_\text{y} \}\}$}&  & \checkmark &\checkmark \\
\hline

$\Psi_5$ & 
\makecell[l]{
$4\mathbb{E}\{|b_\text{x}|^2 \}\mathbb{E}^2\{|a_\text{x}|^2 \}+4\mathbb{E}\{|b_\text{x}|^2 \}|\mathbb{E}\{a_\text{x}^{\ast}a_\text{y} \}|^2+\mathbb{E}\{|b_\text{y}|^2 \}\mathbb{E}\{|a_\text{x}|^2 \}\mathbb{E}\{|a_\text{y}|^2 \}+\mathbb{E}\{|b_\text{x}|^2 \}\mathbb{E}^2\{|a_\text{y}|^2 \}$\\
$+2\Re\{2\mathbb{E}\{b_\text{x}b_\text{y}^{\ast} \}\mathbb{E}\{|a_\text{x}|^2 \}\mathbb{E}\{a_\text{x}^{\ast}a_\text{y} \}+\mathbb{E}\{b_\text{x}b_\text{y}^{\ast} \}\mathbb{E}\{|a_\text{y}|^2 \}\mathbb{E}\{a_\text{x}^{\ast}a_\text{y}\}\}$
}&  & \checkmark & \checkmark\\
\hline
$\Psi_6$ & 
\makecell[l]{
$4\mathbb{E}\{|b_\text{x}|^2 \}|\mathbb{E}\{a_\text{x}^2 \}|^2+4\mathbb{E}\{|b_\text{x}|^2 \}|\mathbb{E}\{a_\text{x}a_\text{y} \}|^2+\mathbb{E}\{|b_\text{y}|^2 \}|\mathbb{E}\{a_\text{x}a_\text{y} \}|^2+\mathbb{E}\{|b_\text{x}|^2 \}|\mathbb{E}\{a_\text{y}^2 \}|^2$\\
$+2\Re\{2\mathbb{E}\{b_\text{x}b_\text{y}^{\ast} \}\mathbb{E}^{\ast}\{a_\text{x}^2 \}\mathbb{E}\{a_\text{x}a_\text{y} \}+\mathbb{E}\{b_\text{x}b_\text{y}^{\ast} \}\mathbb{E}\{a_\text{y}^2 \}\mathbb{E}^{\ast}\{a_\text{x}a_\text{y} \}\}$
}&  & \checkmark & \checkmark\\
\hline
$\Psi_7 $ &  			
\makecell[l]{
$4\mathbb{E}\{|b_\text{x}|^2 \}\mathbb{E}\{|a_\text{x}|^4 \}-8\mathbb{E}\{|b_\text{x}|^2 \}\mathbb{E}^2\{|a_\text{x}|^2 \}-4\mathbb{E}\{|b_\text{x}|^2 \}|\mathbb{E}\{a_\text{x}^2 \}|^2-\mathbb{E}\{|b_\text{y}|^2 \}|\mathbb{E}\{a_\text{x}a^{\ast}_\text{y} \}|^2-\mathbb{E}\{|b_\text{y}|^2 \}|\mathbb{E}\{a_\text{x}a_\text{y} \}|^2$\\
$+\mathbb{E}\{|b_\text{x}|^2 \}\mathbb{E}\{|a_\text{y}|^4 \}-2\mathbb{E}\{|b_\text{x}|^2 \}\mathbb{E}^2\{|a_\text{y}|^2 \}-\mathbb{E}\{|b_\text{x}|^2 \}|\mathbb{E}\{a_\text{y}^2 \}|^2-2\mathbb{E}\{|b_\text{x}|^2 \}|\mathbb{E}\{a^{\ast}_\text{x}a_\text{y} \}|^2-2\mathbb{E}\{|b_\text{x}|^2 \}|\mathbb{E}\{a_\text{x}a_\text{y}^{\ast} \}|^2$\\
$-4\mathbb{E}\{|b_\text{x}|^2 \}|\mathbb{E}\{a_\text{x}a_\text{y} \}|^2+4\mathbb{E}\{|b_\text{x}|^2 \}\mathbb{E}\{|a_\text{x}|^2|a_\text{y}|^2 \}+\mathbb{E}\{|b_\text{y}|^2 \}\mathbb{E}\{|a_\text{x}|^2|a_\text{y}|^2 \}-4\mathbb{E}\{|b_\text{x}|^2 \}\mathbb{E}\{|a_\text{x}|^2\}\mathbb{E}\{|a_\text{y}|^2 \}$\\
$-4\mathbb{E}\{|b_\text{y}|^2 \}\mathbb{E}\{|a_\text{x}|^2\}\mathbb{E}\{|a_\text{y}|^2 \}+2\Re\{-2\mathbb{E}\{b_\text{x}b^{\ast}_\text{y} \}\mathbb{E}\{a_\text{x}a_\text{y} \}\mathbb{E}^{\ast}\{a_\text{x}^2 \}-2\mathbb{E}\{b_\text{x}b^{\ast}_\text{y} \}\mathbb{E}\{|a_\text{x}|^2 \}\mathbb{E}\{a_\text{x}^{\ast}a_\text{y} \}$\\
$-\mathbb{E}\{b_\text{x}b^{\ast}_\text{y} \}\mathbb{E}\{|a_\text{y}|^2 \}\mathbb{E}\{a_\text{x}^{\ast}a_\text{y} \}-\mathbb{E}\{b_\text{x}b^{\ast}_\text{y} \}\mathbb{E}\{a_\text{y}^2 \}\mathbb{E}^{\ast}\{a_\text{x}a_\text{y} \}\}$
}&  & \checkmark & \checkmark\\
\hline

$\Lambda_7$ & 
\makecell[l]{
$\mathbb{E}\{|b_\text{x}|^2 \}\mathbb{E}^2\{|a_\text{x}|^2 \}+\mathbb{E}\{|b_\text{y}|^2 \}\mathbb{E}\{|a_\text{x}|^2 \}\mathbb{E}\{|a_\text{y}|^2 \}+2\Re\{\mathbb{E}\{b_\text{x}^{\ast}b_\text{y} \}\mathbb{E}\{|a_\text{x}|^2 \}\mathbb{E}\{a_\text{x}a_\text{y}^{\ast} \}\}$
}&  & \checkmark & \checkmark\\
\hline
$\Lambda_8$ & 
\makecell[l]{
$\mathbb{E}\{|b_\text{x}|^2 \}\mathbb{E}^2\{|a_\text{x}|^2 \}+\mathbb{E}\{|b_\text{y}|^2 \}|\mathbb{E}\{a_\text{x}^{\ast}a_\text{y} \}|^2+2\Re\{\mathbb{E}\{b_\text{x}^{\ast}b_\text{y} \}\mathbb{E}\{|a_\text{x}|^2 \}\mathbb{E}\{a_\text{x}a_\text{y}^{\ast} \}\}$
}&  & \checkmark & \checkmark\\
\hline
$\Lambda_9$ & 
\makecell[l]{
$\mathbb{E}\{|b_\text{x}|^2 \}\mathbb{E}\{|a_\text{x}|^4 \}-2\mathbb{E}\{|b_\text{x}|^2 \}\mathbb{E}^2\{|a_\text{x}|^2 \}-\mathbb{E}\{|b_\text{x}|^2 \}|\mathbb{E}\{a_\text{x}^2 \}|^2-\mathbb{E}\{|b_\text{y}|^2 \}|\mathbb{E}\{a_\text{x}a_\text{y} \}|^2-\mathbb{E}\{|b_\text{y}|^2 \}|\mathbb{E}\{a_\text{x}^{\ast}a_\text{y} \}|^2$\\
$+\mathbb{E}\{|b_\text{y}|^2 \}\mathbb{E}\{|a_\text{x}|^2|a_\text{y}|^2 \}-\mathbb{E}\{|b_\text{y}|^2 \}\mathbb{E}\{|a_\text{x}|^2 \}\mathbb{E}\{|a_\text{y}|^2 \}+2\Re\{-\mathbb{E}\{b_\text{x}^{\ast}b_\text{y} \}\mathbb{E}\{a_\text{x}^2 \}\mathbb{E}^{\ast}\{a_\text{x}a_\text{y} \}-\mathbb{E}\{b_\text{x}^{\ast}b_\text{y} \}\mathbb{E}\{|a_\text{x}|^2 \}\mathbb{E}\{a_\text{x}a_\text{y}^{\ast} \}\}$
}&  &\checkmark & \checkmark\\
\hline
\hline
\end{tabular}
		}
		\label{tab:two}
\end{table*}

\begin{table*}[!htbp]
		\centering
		\caption{Expression for the terms used in \eqref{SCI_PSD}, \eqref{XCI_PSD} and \eqref{MCI_PSD}}
  \renewcommand\arraystretch{2}
		\changed{
        \scalebox{0.96}{
		\begin{tabular}{c|c}
\hline
\hline
Term & Expression  \\
\hline
\hline
$\chi^{(1)}_{\text{SCI}}$ & 
\makecell[l]{
$\int_{-R_s/2}^{R_s/2}\int_{-R_s/2}^{R_s/2}|P_{\text{COI}}(f_1)|^2|P_{\text{COI}}(f_2)|^2|P_{\text{COI}}(f-f_1+f_2)|^2|\mu(f_1,f_2,f)|^2 df_1df_2$
} \\
\hline
$\chi^{(2)}_{\text{SCI}}$ & 
\makecell[l]{
$\int_{-R_s/2}^{R_s/2}\int_{-R_s/2}^{R_s/2}|P_{\text{COI}}(f_1)|^2|P_{\text{COI}}(f_2)|^2|P_{\text{COI}}(f-f_1+f_2)|^2\mu(f_1,f_2,f)\mu^*(f_1,f_1-f_2-f,f) df_1df_2$
} \\
\hline
$\chi^{(3)}_{\text{SCI}}$ & 
\makecell[l]{
$|P_{\text{COI}}(f)|^2\int_{-R_s/2}^{R_s/2}\int_{-R_s/2}^{R_s/2}|P_{\text{COI}}(f_1)|^2|P_{\text{COI}}(f_2)|^2\mu(f_1,-f,f)\mu^{\ast}(f_2,-f,f) df_1df_2$
}  \\
\hline
$\chi^{(4)}_{\text{SCI}}$ & 
\makecell[l]{
$\int_{-R_s/2}^{R_s/2}\int_{-R_s/2}^{R_s/2}\int_{-R_s/2}^{R_s/2}P_{\text{COI}}(f_1)P_{\text{COI}}^{\ast}(f_2)P_{\text{COI}}(f-f_1+f_2)P_{\text{COI}}^{\ast}(f_1-f_2)P_{\text{COI}}(f'_2)P_{\text{COI}}^{\ast}(f-f_1+f_2+f'_2)\mu(f_1,f_2,f)$\\
$\cdot\mu^{\ast}(f_1-f_2,f'_2,f) df_1df_2df'_2$
}  \\
\hline
$\chi^{(5)}_{\text{SCI}}$ & 
\makecell[l]{
$\int_{-R_s/2}^{R_s/2}\int_{-R_s/2}^{R_s/2}\int_{-R_s/2}^{R_s/2}P_{\text{COI}}(f_1)P_{\text{COI}}^{\ast}(f_2)P_{\text{COI}}(f-f_1+f_2)P_{\text{COI}}(f_2-f_1)P_{\text{COI}}^{\ast}(f'_2)P_{\text{COI}}^{\ast}(f-f_1+f_2-f'_2)$\\
$\cdot\mu(f_1,f_2,f)\mu^{\ast}(f'_2,f_2-f_1,f) df_1df_2df'_2$
}\\
\hline
$\chi^{(6)}_{\text{SCI}}$ & 
\makecell[l]{
$\int_{-R_s/2}^{R_s/2}\int_{-R_s/2}^{R_s/2}\int_{-R_s/2}^{R_s/2}P_{\text{COI}}(f_1)P_{\text{COI}}^{\ast}(f_2)P_{\text{COI}}(f-f_1+f_2)P_{\text{COI}}^{\ast}(f+f_2)P_{\text{COI}}^{\ast}(f'_2)P_{\text{COI}}(f_2+f'_2)\mu(f_1,f_2,f)$\\
$\cdot\mu^{\ast}(f'_2,-f-f_2,f) df_1df_2df'_2$
}\\
\hline
$\chi^{(7)}_{\text{SCI}}$ & 
\makecell[l]{
$P_{\text{COI}}(f)\int_{-R_s/2}^{R_s/2}\int_{-R_s/2}^{R_s/2}\int_{-R_s/2}^{R_s/2}|P_{\text{COI}}(f_1)|^2P_{\text{COI}}^{\ast}(f_2)P_{\text{COI}}(f'_2)P_{\text{COI}}^{\ast}(f-f_2+f'_2)\mu(f_1,-f,f)\mu^{\ast}(f_2,f'_2,f) df_1df_2df'_2$\\
} \\
\hline
$\chi^{(8)}_{\text{SCI}}$ & 
\makecell[l]{
$\int_{-R_s/2}^{R_s/2}\int_{-R_s/2}^{R_s/2}\int_{-R_s/2}^{R_s/2}|P_{\text{COI}}(f_1)|^2P_{\text{COI}}^{\ast}(f_2)P_{\text{COI}}(f-f_1+f_2)P_{\text{COI}}(f'_2)P_{\text{COI}}^{\ast}(f-f_1+f'_2)\mu(f_1,f_2,f)\mu^{\ast}(f_1,f'_2,f) df_1df_2df'_2$\\
} \\
\hline
$\chi^{(9)}_{\text{SCI}}$ & 
\makecell[l]{
$\int_{-R_s/2}^{R_s/2}\int_{-R_s/2}^{R_s/2}\int_{-R_s/2}^{R_s/2}|P_{\text{COI}}(f_1)|^2P_{\text{COI}}^{\ast}(f_2)P_{\text{COI}}(f-f_1+f_2)P_{\text{COI}}^{\ast}(f'_2)P_{\text{COI}}^{\ast}(f-f_1-f'_2)\mu(f_1,f_2,f)\mu^{\ast}(f'_2,-f_1,f) df_1df_2df'_2$\\
} \\
\hline
$\chi^{(10)}_{\text{SCI}}$ & 
\makecell[l]{
$\int_{-R_s/2}^{R_s/2}\int_{-R_s/2}^{R_s/2}\int_{-R_s/2}^{R_s/2}P_{\text{COI}}(f_1)|P_{\text{COI}}(f_2)|^2P_{\text{COI}}(f-f_1+f_2)P_{\text{COI}}^{\ast}(f'_2)P_{\text{COI}}^{\ast}(f+f_2-f'_2)\mu(f_1,f_2,f)\mu^{\ast}(f'_2,f_2,f) df_1df_2df'_2$\\
} \\
\hline
$\chi^{(11)}_{\text{SCI}}$ & 
\makecell[l]{
$\int_{-R_s/2}^{R_s/2}\int_{-R_s/2}^{R_s/2}\int_{-R_s/2}^{R_s/2}\int_{-R_s/2}^{R_s/2}P_{\text{COI}}(f_1)P_{\text{COI}}^{\ast}(f_2)P_{\text{COI}}(f-f_1+f_2)P_{\text{COI}}^{\ast}(f'_2)P_{\text{COI}}(f_4)P_{\text{COI}}^{\ast}(f-f'_2+f_4)$\\
$\cdot\mu(f_1,f_2,f)\mu^{\ast}(f'_2,f_4,f) df_1df_2df'_1df'_2$\\
}\\
\hline 

$\chi^{(1)}_{\text{XCI,X1}}$ & 
\makecell[l]{
$\int_{-R_s/2}^{R_s/2}\int_{f_\text{c}^h-R_s/2}^{f_\text{c}^h+R_s/2}|P_{\text{COI}}(f_1)|^2|P_{\text{INT}}(f_2)|^2|P_{\text{INT}}(f-f_1+f_2)|^2|\mu(f_1,f_2,f)|^2 df_1df_2$
} \\
\hline
$\chi^{(2)}_{\text{XCI,X1}}$ & 
\makecell[l]{
$\int_{-R_s/2}^{R_s/2}\int_{f_\text{c}^h-R_s/2}^{f_\text{c}^h+R_s/2}|P_{\text{COI}}(f_1)|^2P_{\text{INT}}(f_2)P_{\text{INT}}^{\ast}(f-f_1-f_2+2f_\text{c}^h)P_{\text{INT}}(f-f_1+f_2)P_{\text{INT}}^{\ast}(2f_\text{c}^h-f_2)\mu(f_1,f_2,f)$\\
$\cdot\mu^{\ast}(f_1,f_1-f_2-f+2f_\text{c}^h,f) df_1df_2$
} \\
\hline
$\chi^{(3)}_{\text{XCI,X1}} $ &  			
\makecell[l]{
$\int_{-R_s/2}^{R_s/2}\int_{f_\text{c}^h-R_s/2}^{f_\text{c}^h+R_s/2}\int_{f_\text{c}^h-R_s/2}^{f_\text{c}^h+R_s/2}|P_{\text{COI}}(f_1)|^2P_{\text{INT}}(f_2)P_{\text{INT}}^{\ast}(f'_2)P_{\text{INT}}(f-f_1+f_2)P_{\text{INT}}^{\ast}(f-f_1+f'_2)\mu(f_1,f_2,f)\mu^{\ast}(f_1,f'_2,f) df_1df_2df'_2$
} \\ 
\hline
$\chi^{(1)}_{\text{XCI,X2}}$ & 
\makecell[l]{
$\int_{f_\text{c}^h-R_s/2}^{f_\text{c}^h+R_s/2}\int_{-R_s/2}^{R_s/2}|P_{\text{INT}}(f_1)|^2|P_{\text{COI}}(f_2)|^2|P_{\text{COI}}(f-f_1+f_2)|^2|\mu(f_1,f_2,f)|^2 df_1df_2$
} \\
\hline
$\chi^{(2)}_{\text{XCI,X2}}$ & 
\makecell[l]{
$\int_{f_\text{c}^h-R_s/2}^{f_\text{c}^h+R_s/2}\int_{-R_s/2}^{R_s/2}|P_{\text{INT}}(f_1)|^2|P_{\text{COI}}(f_2)|^2|P_{\text{COI}}(f-f_1+f_2)|^2\mu(f_1,f_2,f)\mu^{\ast}(f_1,f_1-f_2-f,f) df_1df_2$
} \\
\hline
$\chi^{(3)}_{\text{XCI,X2}} $ &  			
\makecell[l]{
$\int_{f_\text{c}^h-R_s/2}^{f_\text{c}^h+R_s/2}\int_{-R_s/2}^{R_s/2}\int_{-R_s/2}^{R_s/2}|P_{\text{INT}}(f_1)|^2P_{\text{COI}}(f_2)P_{\text{COI}}^{\ast}(f'_2)P_{\text{COI}}(f-f_1+f_2)P_{\text{COI}}^{\ast}(f-f_1+f'_2)\mu(f_1,f_2,f)\mu^{\ast}(f_1,f'_2,f) df_1df_2df'_2$
}\\ 
\hline
$\chi^{(1)}_{\text{XCI,X3}}$ & 
\makecell[l]{
$\int_{-R_s/2}^{R_s/2}\int_{f_\text{c}^h-R_s/2}^{f_\text{c}^h+R_s/2}|P_{\text{COI}}(f_1)|^2|P_{\text{INT}}(f_2)|^2|P_{\text{COI}}(f-f_1+f_2)|^2|\mu(f_1,f_2,f)|^2 df_1df_2$
} \\
\hline
$\chi^{(2)}_{\text{XCI,X3}}$ & 
\makecell[l]{
$\int_{-R_s/2}^{R_s/2}\int_{f_\text{c}^h-R_s/2}^{f_\text{c}^h+R_s/2}|P_{\text{COI}}(f_1)|^2|P_{\text{INT}}(f_2)|^2|P_{\text{COI}}(f-f_1+f_2)|^2\mu(f_1,f_2,f)\mu^*(f-f_1+f_2,f_2,f) df_1df_2$
}  \\
\hline
$\chi^{(3)}_{\text{XCI,X3}}$ & 
\makecell[l]{
$\int_{-R_s/2}^{R_s/2}\int_{f_\text{c}^h-R_s/2}^{f_\text{c}^h+R_s/2}\int_{-R_s/2}^{R_s/2}P_{\text{COI}}(f_1)|P_{\text{INT}}(f_2)|^2P_{\text{COI}}(f-f_1+f_2)P_{\text{COI}}^{\ast}(f'_2)P_{\text{COI}}^{\ast}(f-f'_2+f_2)\mu(f_1,f_2,f)\mu^{\ast}(f'_2,f_2,f) df_1df_2df'_2$\\
}\\
\hline

$\chi^{(1)}_{\text{MCI,M1}}$ & 
\makecell[l]{
$\int_{f_\text{c}^h-R_s/2}^{f_\text{c}^h+R_s/2}\int_{f_\text{c}^h-R_s/2}^{f_\text{c}^h+R_s/2}|P_{\text{INT}_1}(f_1)|^2|P_{\text{INT}_h}(f_2)|^2|P_{\text{INT}_h}(f-f_1+f_2)|^2|\mu(f_1,f_2,f)|^2 df_1df_2$   (with $h \in$ M1)
}\\
\hline
$\chi^{(2)}_{\text{MCI,M1}}$ & 
\makecell[l]{
$\int_{f_\text{c}^h-R_s/2}^{f_\text{c}^h+R_s/2}\int_{f_\text{c}^h-R_s/2}^{f_\text{c}^h+R_s/2}|P_{\text{INT}_1}(f_1)|^2P_{\text{INT}_h}(f_2)P_{\text{INT}_h}^{\ast}(f-f_1-f_2+2f_\text{c}^h)P_{\text{INT}_h}(f-f_1+f_2)P_{\text{INT}_h}^{\ast}(2f_\text{c}^h-f_2)\mu(f_1,f_2,f)$\\
$\cdot\mu^{\ast}(f_1,f_1-f_2-f+2f_\text{c}^h,f) df_1df_2$  (with $h \in$ M1)
}\\
\hline
$\chi^{(3)}_{\text{MCI,M1}} $ &  			
\makecell[l]{
$\int_{f_\text{c}^h-R_s/2}^{f_\text{c}^h+R_s/2}\int_{f_\text{c}^h-R_s/2}^{f_\text{c}^h+R_s/2}\int_{f_\text{c}^h-R_s/2}^{f_\text{c}^h+R_s/2}|P_{\text{INT}_1}(f_1)|^2P_{\text{INT}_h}(f_2)P_{\text{INT}_h}^{\ast}(f'_2)P_{\text{INT}_h}(f-f_1+f_2)P_{\text{INT}_h}^{\ast}(f-f_1+f'_2)$\\
$\cdot\mu(f_1,f_2,f)\mu^{\ast}(f_1,f'_2,f) df_1df_2df'_2$ 
(with $h \in$ M1)
}\\
\hline
$\chi^{(1)}_{\text{MCI,M2}}$ & 
\makecell[l]{
$\int_{f_\text{c}^h-R_s/2}^{f_\text{c}^h+R_s/2}\int_{f_\text{c}^h-R_s/2}^{f_\text{c}^h+R_s/2}|P_{\text{INT}_1}(f_1)|^2|P_{\text{INT}_h}(f_2)|^2|P_{\text{INT}_h}(f-f_1+f_2)|^2|\mu(f_1,f_2,f)|^2 df_1df_2$  (with $h \in$ M2)
}\\
\hline
$\chi^{(2)}_{\text{MCI,M2}}$ & 
\makecell[l]{
$\int_{f_\text{c}^h-R_s/2}^{f_\text{c}^h+R_s/2}\int_{f_\text{c}^h-R_s/2}^{f_\text{c}^h+R_s/2}|P_{\text{INT}_1}(f_1)|^2P_{\text{INT}_h}(f_2)P_{\text{INT}_h}^{\ast}(f-f_1-f_2+2f_\text{c}^h)P_{\text{INT}_h}(f-f_1+f_2)P^{\ast}_{\text{INT}_h}(2f^h_\text{c}-f_2)$\\
$\cdot\mu(f_1,f_2,f)\mu^{\ast}(f_1,f_1-f_2-f+2f_\text{c}^h,f) df_1df_2$  (with $h \in$ M2)
}\\
\hline
$\chi^{(3)}_{\text{MCI,M2}} $ &  			
\makecell[l]{
$\int_{f_\text{c}^h-R_s/2}^{f_\text{c}^h+R_s/2}\int_{f_\text{c}^h-R_s/2}^{f_\text{c}^h+R_s/2}\int_{f_\text{c}^h-R_s/2}^{f_\text{c}^h+R_s/2}|P_{\text{INT}_1}(f_1)|^2P_{\text{INT}_h}(f_2)P_{\text{INT}_h}^{\ast}(f'_2)P_{\text{INT}_h}(f-f_1+f_2)P_{\text{INT}_h}^{\ast}(f-f_1+f'_2)$\\
$\cdot\mu(f_1,f_2,f)\mu^{\ast}(f_1,f'_2,f) df_1df_2df'_2$  (with $h \in$ M2)
}\\
\hline
$\chi^{(1)}_{\text{MCI,M3}}$ & 
\makecell[l]{
$\int_{f_\text{c}^h-R_s/2}^{f_\text{c}^h+R_s/2}\int_{f_\text{c}^h-R_s/2}^{f_\text{c}^h+R_s/2}|P_{\text{INT}_h}(f_1)|^2|P_{\text{INT}_h}(f_2)|^2|P_{\text{INT}_h'}(f-f_1+f_2)|^2|\mu(f_1,f_2,f)|^2 df_1df_2$  (with $h, h' \in$ M3)
}\\
\hline
$\chi^{(2)}_{\text{MCI,M3}}$ & 
\makecell[l]{
$\int_{f_\text{c}^h-R_s/2}^{f_\text{c}^h+R_s/2}\int_{f_\text{c}^h-R_s/2}^{f_\text{c}^h+R_s/2}|P_{\text{INT}_h}(f_1)|^2|P_{\text{INT}_h}(f_2)|^2|P_{\text{INT}_h'}(f-f_1+f_2)|^2\mu(f_1,f_2,f)\mu(f-f_1+f_2,f_2,f) df_1df_2$  (with $ h, h' \in$ M3)
}\\
\hline
$\chi^{(3)}_{\text{MCI,M3}}$ & 
\makecell[l]{
$\int_{f_\text{c}^h-R_s/2}^{f_\text{c}^h+R_s/2}\int_{f_\text{c}^h-R_s/2}^{f_\text{c}^h+R_s/2}\int_{f_\text{c}^h-R_s/2}^{f_\text{c}^h+R_s/2}P_{\text{INT}_h}(f_1)|P_{\text{INT}_h}(f_2)|^2P_{\text{INT}_h'}(f-f_1+f_2)P_{\text{INT}_h}^{\ast}(f'_2)P_{\text{INT}_h'}^{\ast}(f+f_2-f'_2)$\\
$\cdot\mu(f_1,f_2,f)\mu^{\ast}(f'_2,f_2,f) df_1df_2df'_2$  (with  $h, h' \in$ M3)\\
}\\

\hline
\hline
\multicolumn{2}{p{18cm}}{
The value range of $h, h'$ (i.e., M1, M2 and M3) is shown in Appendix.~\ref{sec: appendix_B}.}
\end{tabular}
		          }
                }
		\label{tab:three}
  \vspace{-1em}
 \end{table*}

\begin{theorem}\label{theorem1}
For a generic aperiodic transmitted signal and a WDM fiber transmission system like the one in Fig.~\ref{fig:System_Model} (a), the coefficient of h-th INT channel $\eta_{\text{ss,XCI}}^{\text{x},h}$ for the x polarization can be written as 
\vspace{-0.4em}
\changed{ \begin{align}\label{XCI}
\eta_{\text{ss,XCI}}^{\text{x},h}&\triangleq \frac{\sigma^2_{\text{XCI,x},h}}{P^3} = P^{-3}\int_{-\infty}^{\infty}S^{\text{x},h}_{\text{XCI}}(f,N_s,L_s)|P(f)|^2df,
\end{align}
where the XCI PSD $S^{\text{x},h}_{\text{XCI}}(f,N_s,L_s)$ can be written as
\begin{align}\label{XCI_PSD}
\resizebox{1\hsize}{!}{$\begin{aligned}
\begin{split}
    S^{\text{x},h}_{\text{XCI}}(f,N_s,L_s)& = \left(\frac{8}{9}\right)^2\gamma^2\Bigg[  R_s^3[\Phi_4\chi^{(1)}_{\text{XCI,X1}}(f)+\Phi_5\chi^{(2)}_{\text{XCI,X1}}(f)]\\
     &+R_s^2\Phi_6\chi^{(3)}_{\text{XCI,X1}}(f)
     +R_s^3[\Psi_5\chi^{(1)}_{\text{XCI,X2}}(f)+\Psi_6\chi^{(2)}_{\text{XCI,X2}}(f)]\\
     &+R_s^2\Psi_7\chi^{(3)}_{\text{XCI,X2}}(f)
    +R_s^3[\Lambda_7\chi^{(1)}_{\text{XCI,X3}}(f)+\Lambda_8\chi^{(2)}_{\text{XCI,X3}}(f)]\\
     &+R_s^2\Lambda_9\chi^{(3)}_{\text{XCI,X3}}(f)\Bigg]
     +\bar{\eta}^{\text{x},h}_{\text{ss,SCI}},
\end{split}
\end{aligned}$}
\end{align}}where the XCI coefficients are given in Table~\ref{tab:two} and Table~\ref{tab:three}, and $\bar{\eta}^{\text{x},h}_{\text{ss,SCI}}$ is obtained using $\eta^{\text{x}}_{\text{ss,SCI}}$ after swapping $a_\text{x/y}\rightarrow b_{\text{x/y}}$ for the coefficients in Table~\ref{tab:two}\footnote{\changed{All sixth-order correlations depend on the probability of occurrence of the constellation points as Thus, the effect of probabilistic shaping on the NLI can also be captured by the model in this paper.}}, and by changing the integration regions $[R_s/2,-R_s/2] \rightarrow [f_\text{c}^h+R_s/2, f_\text{c}^h-R_s/2]$ for the coefficients in Table~\ref{tab:three}.
\end{theorem}

\begin{proof}
See Appendix~\ref{sec: appendix_A}.
\end{proof}

In Theorem~\ref{theorem1}, the coefficients $\Phi_k, \Psi_k, \Lambda_k, k=4,5,...,9$ in \eqref{XCI} represent modulation-dependent intra- and cross- polarization moments of the DP-4D modulation formats. 
The terms \changed{$\chi^{(k)}_{\text{XCI,X1/X2/X3}}, k=1,2,3$} are integrals related to the channel parameters. 
In addition, the XCI power coefficient can be obtained by summing the x and y components, i.e., $\eta_{\text{ss,XCI}}=\eta_{\text{ss,XCI}}^{\text{x}}+ \eta_{\text{ss,XCI}}^{\text{y}}$ where the $\eta_{\text{ss,XCI}}^{\text{y}}$ can be obtained from \eqref{XCI_tot} by swapping the polarization labels $\text{x}\rightarrow \text{y}$ and $\text{y}\rightarrow \text{x}$. Note that \eqref{XCI} is  valid under the following assumptions:
\begin{itemize}
    \item the sequence of DP-4D transmitted symbols $\boldsymbol{a}_n$ for $n\in \mathbb{Z}$ are independent identically distributed (i.i.d.),
    \item the transmitted pulse $p(t)$ has a rectangular (or quasi-rectangular) spectrum, and
    \item \changed{a first-order regular perturbation (RP) framework in the $\gamma$ coefficient for the solution of the Manakov equation.}
\end{itemize}

\subsection{MCI contribution}
Generally, MCI is always thought as weaker than SCI and XCI, especially in high dispersion fibers, as investigated in \cite{Pogg-invited2012}. This is due to the fact that the higher the dispersion, the faster is the (amplitude) decay of the link function $\mu(f_1,f_2,f)$ (see \eqref{mu}, ahead) away from its maximum. Conversely, the lower the dispersion, the slower the decay. Therefore, to accurately predict the nonlinear interference in various scenarios, the MCI contribution was derived following an approach similar to \cite[Appendix~D]{Carena2014}.

As shown in Fig.~\ref{fig:WDM}, the MCI can be divided into four contributions corresponding to the integration islands marked as M0, M1, M2 and M3. The M1 and M2 (yellow and red) regions have a similar structure as XCI in the region X1 (blue), and M3 (green) is similar to the region X3 (purple). Taking the M1 in the domains locating in the second quadrant, parallel to $f_2$ as an example, the triples of M1 can be shown as  $(f_{\text{INT}_{-1}},f_{\text{INT}_1},f_{\text{INT}_1})$ and the triples of X1 can be shown as $(f_{\text{COI}},f_{\text{INT}_1},f_{\text{INT}_1})$. If the $f_{\text{INT}_{-1}}$ is regarded as $f_{\text{COI}}$, M1 has the similar structure of X1. 
Thus, \eqref{XCI} can be used to approximate the contributions of M1, M2 and M3 islands, where the only difference is the integration limits. In particular, for the M0 island, we take the same approach in \cite{Carena2014}, i.e., the contribution of M0 island is produced entirely according to the GN-model, denoted as $\bar{\eta}^x_{\text{M}0}$. The following theorem gives the MCI contributions.

\begin{theorem}\label{Theorem2}
If all channels are assumed to have the same transmitted power and the total number of channel $N_{ch}$ is odd, the $\eta_{\text{ss,MCI}}=\eta_{\text{ss,MCI}}^{\text{x}}+ \eta_{\text{ss,MCI}}^{\text{y}}$ is similar to \eqref{XCI}, which can be expressed as
\vspace{-0.5em}
\changed{\begin{align}\label{MCI}
\eta_{\text{ss,MCI}}^{\text{x}}&\triangleq \frac{\sigma^2_{\text{MCI,x}}}{P^3} = P^{-3}\int_{-\infty}^{\infty}S^{\text{x}}_{\text{MCI}}(f,N_s,L_s)|P(f)|^2df,
\end{align}
where the MCI PSD $S^{\text{x}}_{\text{MCI}}(f,N_s,L_s)$ is given as
\begin{align}\label{MCI_PSD}
   \begin{split}
    S^{\text{x}}_{\text{MCI}}(f,N_s,&L_s) = 2\cdot\left(\frac{8}{9}\right)^2\gamma^2\Bigg[ R_s^3[\Phi_4\chi^{(1)}_{\text{MCI,M1}}(f)\\
    &+\Phi_5\chi^{(2)}_{\text{MCI,M1}}(f)]
    +R_s^2\Phi_6\chi^{(3)}_{\text{MCI,M1}}(f)\\
    &+R_s^3[\Phi_4\chi^{(1)}_{\text{MCI,M2}}(f)+\Phi_5\chi^{(2)}_{\text{MCI,M2}}(f)]\\
    &+R_s^2\Phi_6\chi^{(3)}_{\text{MCI,M2}}(f)
    +R_s^3[\Lambda_7\chi^{(1)}_{\text{MCI,M3}}(f)\\
    &+\Lambda_8\chi^{(2)}_{\text{MCI,M3}}(f)]+R_s^2\Lambda_9\chi^{(3)}_{\text{MCI,M3}}(f)\Bigg]+\bar{\eta}^x_{\text{M}0},
\end{split}
\end{align}}
where the terms $\Phi_k, \Lambda_k, k=4,5,...,9$ are  same with the \eqref{XCI} and shown in Table~\ref{tab:two}. The terms \changed{$\chi^{(k)}_{\text{MCI,M1/M2/M3}}, k = 1,2,3$} are shown in Table~{\ref{tab:three}}.
\end{theorem}

Theorem~\ref{Theorem2} shows the MCI power coefficient for the x component. By swapping the polarization labels $\text{x}$ and $\text{y}$,  $\eta_{\text{ss,XCI}}^{\text{y}}$ can be obtained. A detailed explanation of the integration regions for M1, M2 and M3 is shown in Appendix~\ref{sec: appendix_B}.

\changed{The proposed  NLI model is similar to the EGN model which also consists of integral terms and modulation dependent coefficients. As shown in \cite[Appendix~C]{Carena14}, all the integral terms can be evaluated by a double integral. The modulation dependent coefficients can be can be calculated easily. Therefore, the computational complexity of the proposed model is just that of a double-integral.
}

\ch{\subsection{Discussion of the non-i.i.d. input symbols situation}
The proposed 4D model can be used to predict the NLI power for all DP-4D modulation formats which are assumed that input symbols are independent identically distributed, including ideal probabilistic shaping (ideal infinite-blocklength) and geometrical shaping. However,  when a  distribution matcher (DM) is used to implement probabilistic shaping, which is currently very popular and practical, the generated symbols are non-i.i.d. (see the example in \cite [Fig.~2]{Wu21}). For the case of non-i.i.d. input symbols, \cite{Dar14ISIT} has shown that the transmitting correlated symbols leads to a new correlated term. In addition, several studies have shown that the effective SNR depends on the shaping blocklength in the  probabilistic amplitude shaping (PAS) structure which generates dependent symbols \cite{Amari19, Fehenberger20}. Therefore, the current version of our proposed model have not considered the  non-i.i.d. input symbols situation. 
However, the heuristic methods and  performance metrics  proposed in \cite {Wu21,Cho21} can be applied to evaluate the properties of the symbol energies within a sliding window, and thus  consider the correlated term. It is very interesting to extend the proposed model to the case of non-i.i.d. input symbols, i.e., PAS with finite blocklength,  as a future research direction.
}

\section{Simulation results and analysis}\label{sec: results}
The numerical validation of the model in this work is performed via SSFM simulations, where the optical nonlinearity is kept as the only noise. The simulated multi-span optical system is described in Table~\ref{tab:parameter}. \changed{To verify the reliability of our proposed model, various 4D modulation formats which shown in Table~\ref{tab:con}, are considered in our simulations.}

In this section, we compared 4D modulation formats in terms of $\eta_{\text{ss}}$ to verify the accuracy of the proposed model. To validate  the $\eta_{\text{ss}}$ value, SSFM does not consider the ASE noise, i.e., the ideal EDFA compensates fiber loss without adding optical noise.~In the absence of other noise sources, $\eta_{\text{ss}} = \eta^{\text{x}}_{\text{ss}}+\eta^{\text{y}}_{\text{ss}}$ can be estimated via the received SNR for COI via the relationship
\begin{align}
    \eta^{\text{x}}_{\text{ss}} \approx \frac{P_{\text{x}}}{\text{SNR}_{\text{eff,x}}^{\text{SSFM}} P^3},
\end{align}
where $P_\text{x}$ and $\text{SNR}^{\text{SSFM}}_\text{eff,x}$ are the transmitted power and the effective SNR over the x polarization, respectively. The value of $\text{SNR}_{\text{eff,x}}^{\text{SSFM}}$ is estimated via \cite [eq.~(22)]{Hami2021}
\begin{align}\label{eff,snr}    \text{SNR}_{\text{eff,x}}^{\text{SSFM}}=\frac{\sum_{j=1}^{M}|\bar{y}_j|^2}{\sum_{j=1}^M\mathbb{E}\{ |Y-\bar{y}_j|^2|X=x_j\}},
\end{align}
\ch{in which the $X$ and $Y$ are random variables (RVs), which assumed to be statistically independent, representing the transmitted symbols and received symbols over x polarisation, respectively.}  
\changed{In \eqref{eff,snr}, $M$ is the number of constellation points in 4D and $x_j$ represents the $j$-th constellation point.}  
The variable $\bar{y}_j$ represents conditional mean, i.e., $\bar{y}_j=\mathbb{E}\{ Y|X=x_j\}$, where $\mathbb{E}\{ \cdot \}$ is the statistical expectation.

\begin{table}[!tb]
		\centering
		\caption{System and Fiber Parameters}
		\small
		\renewcommand{\arraystretch}{0.95}
		\scalebox{0.94}{\begin{footnotesize}
\begin{tabular}{c|c|c}
\hline\hline
 & {\textbf{Parameter}} & \textbf{Value}\\
\hline
\multirow{5}{*}{\bf TX parameters} & Symbol rate ($R_s$)  & $45$~GBaud \\
\cline{2-3}
&\multirow{2}{*}{No. of channels} & 1 (Figs.~\ref{fig:P-sn},  \ref{fig:ngmi}(a))\\
\cline{3-3}
& & 9 (Figs.~\ref{fig:XCI}, \ref{fig:diff_const}, \ref{fig:ngmi}(b))\\
\cline{2-3}
& RRC rolloff & 0.01\% \\
\cline{2-3}
& \multirow{2}{*}{\changed{Tx power per channel ($P$)}} & $-20$~dBm (Figs.~\ref{fig:XCI}--\ref{fig:diff_const})\\
\cline{3-3}
& & \changed{$0.5$~dBm (Figs.~\ref{fig:P-sn})}\\
\hline
\multirow{2}{*}{\bf Link parameters}
& Span length & $80$~km \\
\cline{2-3}
& \multirow{1}{*}{No. of spans} & 20 (Fig.~\ref{fig:diff_const})\\
\hline
\hline
\multirow{3}{*}{\textbf{SMF} (Figs.~\ref{fig:P-sn}, \ref{fig:XCI} -- \ref{fig:ngmi})}
& Attenuation coeff. ($\alpha$) & $0.2$~dB/km \\
\cline{2-3}
& Dispersion par. ($D$) & $17$~ps/nm/km\\
\cline{2-3}
& Nonlinear coeff. ($\gamma$) & $1.3$~$(\text{W}\cdot \text{km})^{-1}$\\
\hline
\multirow{3}{*}{\textbf{NZDSF} (Figs.~\ref{fig:XCI})}
& Attenuation coeff. ($\alpha$) & $0.2$~dB/km \\
\cline{2-3}
& Dispersion par. ($D$) &  $3.8$~ps/nm/km\\
\cline{2-3}
& Nonlinear coeff. ($\gamma$) & $1.5$~$(\text{W}\cdot \text{km})^{-1}$\\
\hline
\multirow{3}{*}{\textbf{LDF} (Figs.~\ref{fig:XCI})}
& Attenuation coeff. ($\alpha$) & $0.2$~dB/km \\
\cline{2-3}
& Dispersion par. ($D$) & $-1.8$~ps/nm/km\\
\cline{2-3}
& Nonlinear coeff. ($\gamma$) & $2.2$~$(\text{W}\cdot \text{km})^{-1}$\\
\hline
\hline
\multirow{2}{*}{\textbf{SSFM parameters}}
& Step size & $0.1$~km\\
\cline{2-3}
& Samples per symbol & 4\\
\cline{2-3}
& Noise figure & 5~dB\\
\hline\hline
\end{tabular}
\end{footnotesize}

}
		\label{tab:parameter}
\end{table}

\begin{table}[!tb]
\caption{\changed{Considered modulation formats in the paper.}}
\label{tab:con}
    \centering
  \changed{\scalebox{1}{\begin{tabular}{c|c|c|c}
\hline\hline
$M$ & {\textbf{Const. label}}  & \textbf{Symmetric} & \textbf{Constant modulus}\\
\hline

8 & l4\_8\cite{Karlsson:09} & \checkmark & \checkmark  \\
\hline
\multirow{2}{*}{16} & c4\_16\cite{Karlsson2010} & \ding{56} & \ding{56}\\
\cline{2-4}
& cube4\_16\cite{Welti1974}  & \checkmark & \ding{56}\\
\hline

\multirow{2}{*}{32} & voronoi4\_32\cite{Forney1989} & \ding{56} & \ding{56}\\
\cline{2-4}
& b4\_32\cite{Biglieri1992} & \checkmark & \ding{56}\\
\hline

\multirow{2}{*}{64} & w4\_64\cite{Welti1974} &\ding{56} & \ding{56}\\
\cline{2-4}
 & 4D-PRS64\cite{BinChenJLT2019} & \checkmark & \checkmark\\
\hline

\multirow{3}{*}{128} & l4\_128\cite{Database} & \ding{56} & \ding{56}\\
\cline{2-4}
 & 4D-2A-8PSK7b\cite{Kojima2017JLT} & \checkmark & \checkmark\\
 \cline{2-4}
 & 4D-OS128\cite{2020Bin-OS} & \checkmark & \ding{56}\\
\hline

\multirow{2}{*}{256} & ab4\_256\cite{Eriksson:15} & \checkmark & \ding{56}\\
\cline{2-4}
 & w4\_256\cite{Welti1974} & \checkmark & \ding{56}\\
\hline
512 & sphere\_512\cite{Database} & \checkmark & \ding{56}\\
\hline
1024 & a4\_1024\cite{Database} & \checkmark & \ding{56}\\
\hline
2048 & a4\_2048\cite{Database} & \checkmark & \ding{56}\\
\hline
\multirow{2}{*}{4096} & a4\_4096\cite{Database} & \checkmark & \ding{56}\\
\cline{2-4}
& PM-64QAM & \checkmark & \ding{56}\\





\hline\hline
\end{tabular}}}
\end{table}

\begin{figure*}[!tb]
\centering
      \scalebox{1}{\includegraphics{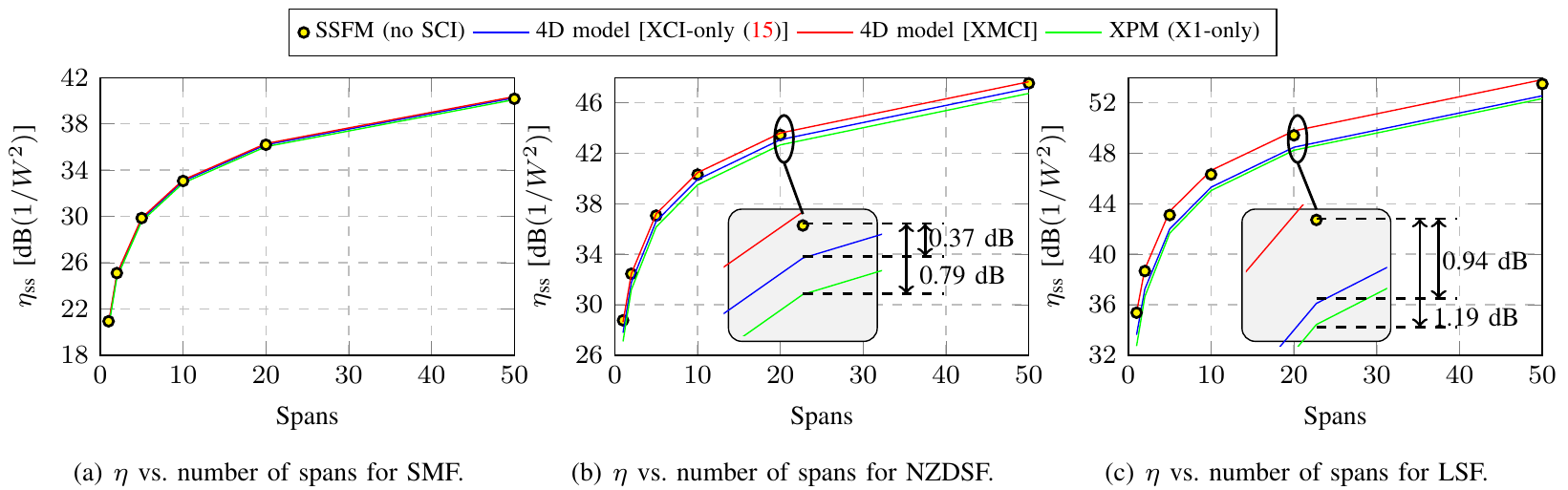}}
\vspace{-0.5em}
\caption{Simulation results of multi-span 9-channel optical fiber transmission system for three fiber types: (a) SMF, (b) NZDSF and (c) LDF. The nonsymmetric constellation voronoi4\_32 \cite{Forney1989} was used for transmission. Blue solid line is $\eta_{\text{ss,XCI}}$ in \eqref{XCI}. Red solid line is $\eta_{\text{ss,XMCI}}$ (i.e., XCI + MCI). Green solid line is the XPM $\eta_{\text{ss,XPM}}$, which is approximately equal to $\eta_{\text{ss,X1}}$ in \eqref{ss,xci}. The marks are SSFM simulations with single-channel nonlinearity (SCI) removed.}
\label{fig:XCI}
\end{figure*}

\changed{The NLI model we consider in this paper was derived under a perturbation theory framework. In addition, as we all known, $\eta_{\text{ss}}$ is a function of several system parameters, albeit independent of the transmitted power.   Therefore, the NLI model can be used  to predict optical communication system performance in the linear and pseudo-linear regimes. Here to validate the accuracy of the proposed NLI model, every channel performed at both low and optimal launch power (-20~dBm and 0.5~dBm, resp.).
}
\begin{figure*}[!tb]
    \centering
      \centering
      \scalebox{1}{\includegraphics{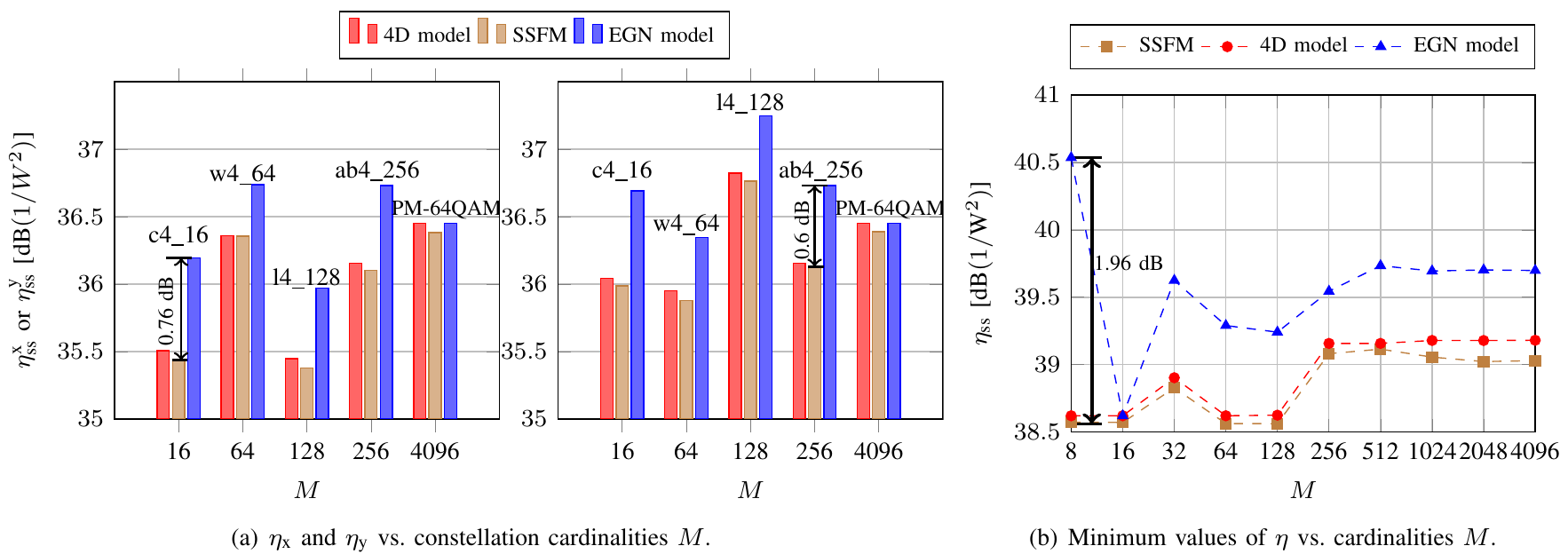}}
      \vspace{-1em}
    \caption{Simulation results of multi-span standard SMF transmission with 9 channels for various DP-4D modulation formats at distance of 1600~km. (a) plots the NLI power coefficient $\eta^\text{x}_{\text{ss}}$ (left) and $\eta^\text{y}_{\text{ss}}$ (right) for different constellation cardinalities $M$: the red bar is the 4D model (including SCI, XCI and MCI), the brown bar is SSFM; the blue bar is EGN model \cite{Carena14}, respectively. (b) plots minimum values of $\eta$ vs. different constellation cardinalities $M$: the blue dashed line is EGN mdoel; the red dashed is 4D model; the brown dashed line is SSFM.}
    \label{fig:diff_const}
    \vspace{-0.5em}
\end{figure*} 

\subsection{Numerical Validation for Multi-Channel Transmission}
In this subsection, we show the NLI power coefficient versus the number of spans for transmitting the nonsymmetric constellation voronoi4\_32 \cite{Forney1989} in three different fiber types, i.e., standard SMF, non-zero dispersion-shifted fiber (NZDSF), low dispersion fiber (LDF), which are listed in Table~\ref{tab:parameter}. To remove SCI, we ran a single-channel simulation and recorded its NLI power coefficient $\eta_{\text{ss,SCI}}$. 
By subtracting $\eta_{\text{ss,SCI}}$ in  total NLI power coefficient $\eta_{\text{ss}}$ of  WDM simulations,  the residual coefficient was estimated as  $\eta_{\text{ss,XMCI}}$  \cite{Poggiolini2015JLT}.

In Fig.~\ref{fig:XCI}, the $\eta_{\text{ss,XPM}}$ is approximately equal to $\eta_{\text{ss,X1}}$ in \eqref{ss,xci}, shown as a green solid line. The blue solid line represents $\eta_{\text{ss,XCI}}$ given by \eqref{XCI}. The marks represent the simulation results which account for all NLI except SCI. Fig.~\ref{fig:XCI} (a) shows that the XCI or XPM is sufficient to present simulated NLI (except SCI) over high accumulated dispersion scenarios for example standard SMF fiber. According to the inset of Fig.~\ref{fig:XCI}~(b), the XPM underestimate the simulated NLI by about 0.79~dB, while the XCI can reduce such error to 0.37~dB. And the inset of Fig.~\ref{fig:XCI}~(c) considered a ultra-low dispersion fiber, showing a wide underestimate error of about 1.19~dB for XPM  and 0.94~dB for XCI. While as shown the red solid line which represents $\eta_{\text{ss,XMCI}}=\eta_{\text{ss,XCI}}+\eta_{\text{ss,MCI}}$, the 4D model with MCI under consideration matches well with the SSFM results for all the three considered fibers including the low and ultra-low dispersion fibers. This suggests that the XCI-only or XPM can not represent all NLI (except SCI), especially in low accumulated dispersion scenarios.

In Fig.~\ref{fig:diff_const} (a), the values of $\eta^{\text{x}}_\text{ss}$ (left) and $\eta^{\text{y}}_\text{ss}$ (right) were estimated using different models for 16-point, 64-point, 128-point, 256-point and 4096-point constellations. We considered (i) the proposed 4D model including SCI, XCI and MCI (red bars), (ii) the EGN model (blue bars), and (iii)  SSFM results (brown bars). For PM-2D modulation formats such as PM-64QAM, our model gives the same result as the EGN model and approximate quite well the SSFM results. For nonsymmetric constellations, the EGN model leads to inaccuracies of up to 0.76~dB for $\eta^{\text{x}}_\text{ss}$ for ``c4\_16" \cite{Karlsson2010}. Even for symmetric constellations, the EGN model also leads to inaccuracies of up to 0.60~dB for $\eta^{\text{y}}_\text{ss}$ in ``ab4\_256"\cite{Eriksson:15}. Such errors between the EGN model results and the SSFM results indicate obvious limitations of the EGN model in predicting the NLI of 4D modulation formats. For all constellations shown, the 4D model has ability to predict NLI of DP-4D modulation formats within acceptable margin of error ($<$ 0.07~dB).

\begin{table}[!tb]
    \centering
    \caption{$\eta$-optimal formats in Fig.~\ref{fig:diff_const} (b)}
    \begin{tabular}{c|c|c}
\hline\hline
{\textbf{Const. label}} & $M$ & ($\eta^{\text{x}}_\text{ss}$,$\eta^{\text{y}}_\text{ss}$) [dB $1/W^2$]\\
\hline
l4\_8 & 8 & (35.6, 35.6)\\
\hline
cube4\_16 & 16 & (35.6, 35.6)\\
\hline
b4\_32 & 32 & (35.9, 35.9)\\
\hline
4D-PRS64 & 64 & (35.6, 35.6)\\
\hline
4D-2A-8PSK7b & 128 & (35.6, 36.6)\\
\hline
w4\_256 & 256 & (35.2, 35.2)\\
\hline
sphere\_512 & 512 & (36.2, 36.2)\\
\hline
a4\_1024 & 1024 & (36.2, 36.2)\\
\hline
a4\_2048 & 2048 & (36.1,36.2)\\
\hline
a4\_4096 & 4096 & (36.3,36.1)\\
\hline\hline
\end{tabular}
    \label{tab:const}
\end{table}

To further validate our proposed 4D model, more 4D modulation formats with different constellation cardinalities against the NLI power coefficient $\eta_{\text{ss}}$ were investigated in Fig.~\ref{fig:diff_const}. Among these, the minimum values of the NLI power coefficient $\eta_{\text{ss}}$ are shown in Fig.~\ref{fig:diff_const}~(b) for each $M$. The corresponding values are also listed in Table~\ref{tab:const} for x- and y- polarization. For all constellations shown, the EGN model overestimates the value of the NLI power coefficient $\eta_{\text{ss}}$ with deviations up to 1.96~dB ($M$=8: l4\_8). 
Conversely, the 4D model is in perfect agreement with the simulation results with the maximum only deviation about 0.15~dB  (for all constellation cardinalities $M$).

\subsection{Analysis in the presence of Signal-ASE Noise beating}
The previous simulation results indicate that the proposed model would be accurate enough to predict the contribution of NLI in short links, where signal-signal noise interaction are predominant. In this section, the effect of signal-ASE noise interaction in the prediction of the effective SNR for general DP-4D modulation formats is evaluated via \eqref{SNR}, including both of the signal-signal interaction and signal-ASE noise interaction.

\begin{figure}[!tb]
 \vspace{-0.0em}
    \centering
       \centering
    {\includegraphics{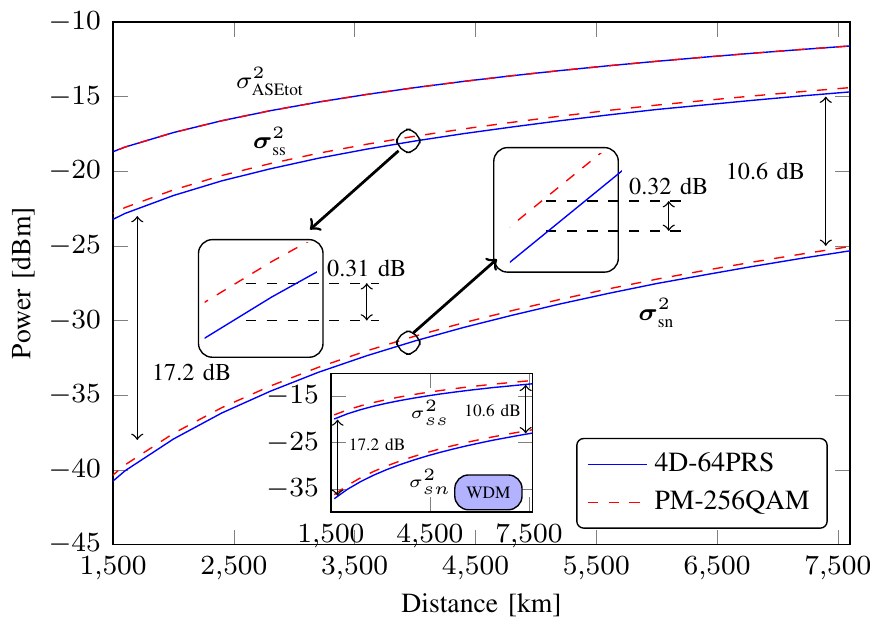}}
    \vspace{-1em}
    \caption{Noise power versus transmission distance at launch power of $P=0.5$~dBm with a single channel. \changed{Inset: Noise power vs. transmission distance with WDM channel.} Noise is shown separately, as  total ASE noise, signal-signal NLI  and signal-ASE noise NLI in \eqref{SNR}.
    }
    \label{fig:P-sn}
     \vspace{-0.5em}
\end{figure}

Fig.~\ref{fig:P-sn} shows the noise powers $\sigma^2_{\text{ASEtot}}$, $\sigma^2_{\text{ss}}$, $\sigma^2_{\text{sn}}$ in \eqref{SNR}, against transmission distance, for two modulation formats: PM-256QAM and 4D-PRS64~\cite{BinChenJLT2019}. The system parameters are shown in Table~\ref{tab:parameter}. For all distance shown, the total ASE noise power is constellation-independent and the NLI contribution of signal-signal and signal-ASE interactions is smaller than the total ASE noise power. Comparing the solid and dashed lines, the dependency of the NLI noise on the modulation format is shown. A 0.3~dB gap can be observed when comparing these two modulation formats at 4000~km. In addition, when these two modulation formats are compared at a distance of $1600$~km, $\sigma^2_{\text{sn}}$ differs from $\sigma^2_{\text{ss}}$ by 17.2~dB. This difference is reduced to 10.6~dB at 7500~km. 
The proportion of $\sigma^2_{\text{sn}}$ in NLI power keeps increasing as the number of fiber span increases. \changed{As shown in the inset of Fig. 5,  the WDM case shows the same conclusion.}  This indicates that the effect of signal-ASE noise NLI can not be fully neglected in long-distance transmission.

\begin{figure}[!tb]
\centering
\scalebox{1}{\includegraphics{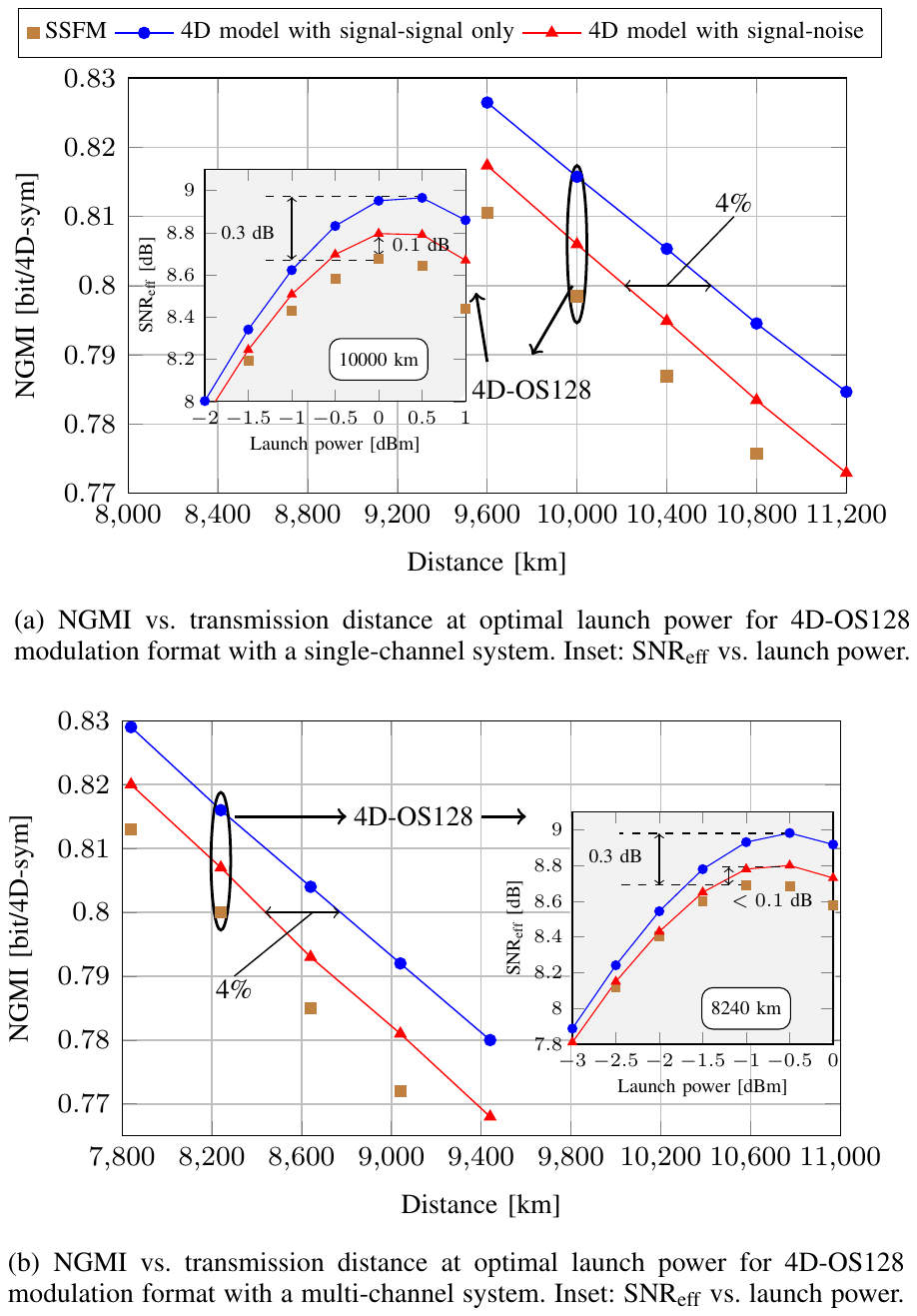}}
\vspace{-2em}
\caption{Transmission performance using 4D-OS128 modulation format. 
    }
\label{fig:ngmi}
\vspace{-1em}
\end{figure}

Fig.~\ref{fig:ngmi} shows the transmission performance estimation in terms of normalized generalized mutual information (NGMI) for the 4D model with signal-noise interactions, using the 4D-OS128 modulation format. In Fig.~\ref{fig:ngmi} (a) and (b), we can observe that  the 4D model with signal-noise interactions can reduce the transmission reach prediction error by 4\% relative to the 4D model with signal-signal interactions only, at NGMI of 0.8 for both single-channel and multi-channel systems.
The prediction accuracy gains come from 
the larger overestimated $\text{SNR}_\text{eff}$ of 0.3~dB for the 4D model with signal-signal interactions only, compared to the SSFM results. As shown in the insets of Fig.~\ref{fig:ngmi}~(a) and (b), the proposed 4D model with signal-noise interactions reduces the SNR  deviation within 0.1~dB for both single-channel and  multi-channel systems, compared to the 4D model with signal-signal interactions only. Therefore, the 4D model with signal-noise provides a better accuracy on  performance prediction  than 4D model with signal-signal interactions only, especially in long-distance transmission.

\section{Conclusions}\label{sec: conclusion}
\ch{An ``ultimate" 4D nonlinear interference model accounting for the intra- and inter-channel nonlinearity for all DP-4D  modulation formats with independent symbols was proposed and validated in detail for multi-channel optical transmission systems.} Unlike the EGN model, which ignores the inter-polarization dependencies, our model makes no assumptions on either the marginal or joint statistics of the two polarization components of the transmitted 4D modulation formats besides being zero mean. 
The proposed model has the ability to predict the SCI, XCI and MCI nonlinear terms---splitted into intra-polarization and cross-polarization terms---for arbitrary DP-4D modulation formats. 
In addition, the proposed model is shown to be accurate for various scenarios, including both high and low dispersion fiber systems. Comparing the induced NLI for  different 4D modulation formats,
the numerical results show that the EGN model overestimates the NLI power up to 1.96~dB, while the proposed 4D NLI model can reduce the   
NLI power estimation error  within 0.15~dB from the SSFM simulation results.

On the other hand, we evaluated the weight of signal-ASE noise interaction in  the prediction of the effective SNR of general DP-4D constellations. Our results show that when signal-ASE noise interactions are considered the accuracy of SNR estimation is improved by about 0.2~dB in single-channel or multi-channel WDM systems.

The proposed 4D NLI model in this work is a powerful analytical tool for the global optimization of 4D modulation formats in the optical fiber channel. \changed{Although the presented results only consider 4D constellations using geometrical shaping, \ch{the proposed NLI model can be also used to predict the NLI power for 4D probabilistic shaping with independent identically distributed input symbols (ideal infinite-blocklength) or finite-blocklength shaping. 
An extension of this work to further extend the proposed model considering  various 4D probabilistic shaping and finite blocklength  will be addressed in a future work.}}

\begin{appendices}
\section{Proof of Theorem~\ref{theorem1}}\label{sec: appendix_A}
To find an analytical expression for NLI power, firstly, a solution to the Manakov equation, which is the fundamental equation of dual-polarization fiber  non-linear dispersive propagation, must be found. We start from the Manakov equation which can be written in time domain as \cite{Marcuse1997}
\begin{equation}
\resizebox{0.99\hsize}{!}{$\begin{aligned}
    \frac{\partial E(t,z)}{\partial z} = -\frac{\alpha}{2}E(t,z)-j\frac{\beta_2}{2}\frac{\partial^{2} E(t,z)}{\partial t^2}+j\frac{8}{9}\gamma |E(t,z)|^2E(t,z),
\end{aligned}$}
\end{equation}
where $\alpha$ is the loss coefficient, $\beta_2$ is dispersion coefficient and $\gamma$ is the nonlinear coefficient. As it is well-known, the Manakov equation has no general closed-form solutions. Like most of the existing NLI power models in the literature \cite{Poggiolini2014, Dar13, Carena14}, the model derived here operates within a first-order perturbative framework. In particular, a frequency-domain first-order RP approach in the $\gamma$ coefficient is performed \cite{Vannucci2002,Johannisson2013}. 
Therefore, the first order RP solution of the Manakov equation after $N_s$ spans is expressed as \cite[eq.~(13)]{GabrieleEntropy2020}
\begin{align}\label{Manakov-solution}
\begin{split}
    \boldsymbol{E}(f,N_s,L_s)&=[E_{\text{x}}(f,N_s,L_s), E_{\text{y}}(f,N_s,L_s)]^{\mathrm{T}}\\
    &=-j\frac{8}{9}\gamma \int^{\infty}_{-\infty}\int^{\infty}_{-\infty} E^{\mathrm{T}}(f_1,0)E^{\ast}(f_2,0)\\
    &\quad E(f-f_1+f_2,0)\mu(f_1,f_2,f,N_s,L_s)df_1df_2.
\end{split}
\end{align}

Due to the lumped amplification and identical spans assumption, $\mu(f_1,f_2,f,N_s,L_s)$ defined in \cite[eq.~19]{Poggiolini2015JLT} can be expressed as
\begin{align}\label{mu}
    \begin{split}
        \mu(f_1,f_2,f,N_s,L_s) \triangleq &\frac{1-e^{-\alpha L_s} e^{j4\pi^2\beta_2(f-f_1)(f_2-f_1)L_s}}{\alpha-j4\pi^2\beta_2(f-f_1)(f_2-f_1)}\\
        &\cdot \sum^{N_s}_{l=1}e^{-j4\pi^2\beta_2(f-f_1)(f_2-f_1)L_s}.
    \end{split}
\end{align}

This formula called the ``link function" is a function of the link parameters and not dependent on characteristics of the launched signal. As shown in this formula, the ``link function" represents the contribution of different NLI fields $(f_1,f_2,f)$ accumulated in different spans so that it weights the generation of NLI and relates to channel parameters.

Under the assumption that the transmitted signal model is periodic with period $T$, where $W$ symbols are transmitted every period $T$, the WDM transmitted signal model can be expressed as \eqref{time-TxModel} and \eqref{Tx} as shown in Sec.~\ref{sec:system}. The Fourier transform of the $E(t,0)$ is given by
\begin{align}\label{general-Tx-model}
\begin{split}
    E(f,0) = \sqrt{\Delta_f}\sum_{h=-(N_{ch}-1)/2}^{(N_{ch}-1)/2}\sum_{k=-\infty}^{+\infty}\boldsymbol{\zeta}_{k,h}\delta(f-f_\text{c}^h-k\Delta_f),
\end{split}
\end{align}
where $\Delta_f = 1/T$ and
\begin{align}
\begin{split}
    \boldsymbol{\zeta}_{k,h}&=[\zeta_{x,k,h},\zeta_{y,k,h}]^{\mathrm{T}}\\
    &=\sqrt{\Delta_f}P(f_\text{c}^h+k\Delta_f)\sum_{n=-(W-1)/2}^{(W-1)/2}\boldsymbol{c}_n e^{-j2\pi\frac{kn}{W}},
\end{split}
\end{align}
is the discrete Fourier transforms of the $h$-th channel transmitted symbol sequence ($\boldsymbol{a}_n$ or $\boldsymbol{b}_n$).

Only considering a COI and an INT channel, the transmitted signal model can be simplified as
\begin{align}\label{Trans-model}
\begin{split}
    E(f,0)=&\sqrt{\Delta_f}\sum_{k=-\infty}^{+\infty}\boldsymbol{\nu}_k\delta(f-k\Delta_f)\\
    &+\sqrt{\Delta_f}\sum_{k=-\infty}^{+\infty}\boldsymbol{\xi}_k\delta(f-f_\text{c}^h-k\Delta_f),
\end{split}
\end{align}
where
\begin{align}\nonumber
\resizebox{0.95\hsize}{!}{$\begin{aligned}
    \boldsymbol{\nu}_k=[\nu_{\text{x},k},\nu_{\text{y},k}]^{\mathrm{T}}=\sqrt{\Delta_f}P(k\Delta_f)\sum_{n=-(W-1)/2}^{(W-1)/2}\boldsymbol{a}_n e^{-j2\pi\frac{kn}{W}},
    \end{aligned}$}
\end{align}
and
\begin{align}\nonumber
\hspace{-0.5em}
    \resizebox{0.95\hsize}{!}{$\begin{aligned}
    \boldsymbol{\xi}_k=[\xi_{\text{x},k},\xi_{\text{y},k}]^{\mathrm{T}}=\sqrt{\Delta_f}P(f_\text{c}^h+k\Delta_f)\sum_{n=-(W-1)/2}^{(W-1)/2}\boldsymbol{b}_n e^{-j2\pi\frac{kn}{W}}.
    \end{aligned}$}
\end{align}
in which $\boldsymbol{a}_n = [a_{\text{x},n},a_{\text{y},n}]^T$ are random vectors transmitted by the COI, complex symbols modulated on two arbitrary orthogonal polarization states x and y, respectively. $\boldsymbol{b}_n = [b_{\text{x},n},b_{\text{y},n}]^T$ is RVs transmitted by an INT channel. 

Substituting the spectrum of the transmitted periodic signal \eqref{Trans-model} in \eqref{Manakov-solution}, 
we obtain the PSD of received NLI for x component,
\begin{align}\label{NLI}
\begin{split}
    & E_\text{x}  (f,N_s,L_s)= -j\frac{8}{9}\gamma\Delta_f^{3/2}\sum^{\infty}_{i=-\infty}\delta(f-i\Delta_f)\\
    &\left[\sum_{\text{S}_i}(\nu_{\text{x},k}\nu^{\ast}_{\text{x},m}\nu_{\text{x},n}+\nu_{\text{y},k}\nu^{\ast}_{\text{y},m}\nu_{\text{x},n})\mu(\text{S}_i,N_s,L_s)\right.\\
    &\left.+\!\sum_{\text{X}1_i}\!(2\nu_{\text{x},k}\xi^{\ast}_{\text{x},m}\xi_{\text{x},n}\!+\nu_{\text{y},k}\xi^{\ast}_{\text{y},m}\xi_{\text{x},n}\!+\nu_{\text{x},k}\xi^{\ast}_{\text{y},m}\xi_{\text{y},n})\!\mu(\text{X}1_i\!,N_s\!,L_s\!)\right.\\
    &\left.\!+\sum_{\text{X}2_i}(2\nu_{\text{x},k}\nu^{\ast}_{\text{x},m}\xi_{\text{x},n}\!+\nu_{\text{y},k}\nu^{\ast}_{\text{y},m}\xi_{\text{x},n}\!+\nu_{\text{x},k}\nu^{\ast}_{\text{y},m}\xi_{\text{y},n})\!\mu\!(\text{X}2_i\!,N_s\!,L_s\!)\right.\\
    &\left.+\sum_{\text{X}3_i}(\nu_{\text{x},k}\xi^{\ast}_{\text{x},m}\nu_{\text{x},n}+\nu_{\text{y},k}\xi^{\ast}_{\text{y},m}\nu_{\text{x},n})\mu(\text{X}3_i,N_s,L_s)\right.\\
    &\left.+\sum_{\text{X}4_i}(\xi_{\text{x},k}\xi^{\ast}_{\text{x},m}\xi_{\text{x},n}+\xi_{\text{y},k}\xi^{\ast}_{\text{y},m}\xi_{\text{x},n})\mu(\text{X}4_i,N_s,L_s)\right.\\
    &\left.+\sum_{\text{X}5_i}(\xi_{\text{x},k}\nu^{\ast}_{\text{x},m}\xi_{\text{x},n}+\xi_{\text{y},k}\nu^{\ast}_{\text{y},m}\xi_{\text{x},n})\mu(\text{X}5_i,N_s,L_s)
    \right],
\end{split}
\end{align}
where
\begin{align}\nonumber
    \begin{split}
        & \text{X}1_i=\text{S}_i=\left\{(k,m,n):(k-m+n)\Delta_f=i\Delta_f \right\}\\
        & \text{X}2_i=\text{X}4_i=\left\{(k,m,n):(k-m+n)\Delta_f+f_\text{c}^h=i\Delta_f\right\}\\
        & \text{X}3_i=\left\{(k,m,n):(k-m+n)\Delta_f-f_\text{c}^h=i\Delta_f\right\}\\
        & \text{X}5_i=\left\{(k,m,n):(k-m+n)\Delta_f+2f_\text{c}^h=i\Delta_f\right\},
    \end{split}
\end{align}
are the integration regions. 

The first summation $S_i$ is SCI, which is dealt with in \cite{GabrieleEntropy2020}. Note that the summation of $\text{X}5_i$ is always zero \cite [Appendix~C]{Carena2014}, when the channels do not overlap, i.e., the INT channel center frequency satisfies the relation of $f_\text{c}^h \ge R_s$.

The PSD of the first order NLI is defined as \cite [eq.~(20)]{GabrieleEntropy2020}
\begin{align}\label{PSD-NLI}
\begin{split}
    \bar{\boldsymbol{S}}(f,N_s,L_s)&=[\bar{S}_\text{x}(f,N_s,L_s),\bar{S}_\text{y}(f,N_s,L_s)]^{\mathrm{T}}\\
    &=[\mathbb{E}\{|E_\text{x}(f,N_s,L_s)|^2\},\mathbb{E}\{|E_\text{y}(f,N_s,L_s)|^2\}],
\end{split}
\end{align}
where $\mathbb{E}\{ \cdot \}$ is the statistical expectation.

Substituting the expression \eqref{NLI} in \eqref{PSD-NLI}, we obtain the PSD of received XCI. 
For the sake of brevity, we just present the detailed derivation of the set $\text{X}1_i$ for the x component. As for the field on the y polarization, it can be found by swapping the subscripts x and y, and the other set can be derived following the same approach. 

For the region $\text{X}1_i$, we have
    \begin{equation}\label{X1}
    \begin{split}
    & S^{\text{x},h}_{\text{XCI},\text{X}1_i}(f,N_s,L_s)= \left(\frac{8}{9}\right)^2\gamma^2\Delta_f^{3}\sum^{\infty}_{i=-\infty}\delta(f-i\Delta_f)\\
    &\cdot \sum_{k,m,n\in \text{X}1_i}\sum_{k',m',n' \in \text{X}1_i}\mu(\text{X}1_i,N_s,L_s)\mu^{\ast}(\text{X}1_i,N_s,L_s)\\
    &\cdot \left[4\mathbb{E}\{\nu_{\text{x},k}\nu^{\ast}_{\text{x},k'} \}\mathbb{E}\{\xi^{\ast}_{\text{x},m}\xi_{\text{x},n}\xi_{\text{x},m'}\xi^{\ast}_{\text{x},n'} \}+ 2\mathbb{E}\{\nu_{\text{x},k}\nu^{\ast}_{\text{y},k'} \}\right.\\
    &\left.\cdot\mathbb{E}\{\xi^{\ast}_{\text{x},m}\xi_{\text{x},n}\xi_{\text{y},m'}\xi^{\ast}_{\text{x},n'} \}+2\mathbb{E}\{\nu_{\text{x},k}\nu^{\ast}_{\text{x},k'} \}\mathbb{E}\{\xi^{\ast}_{\text{x},m}\xi_{\text{x},n}\xi_{\text{y},m'}\xi^{\ast}_{\text{y},n'} \}\right.\\
    &\left.+2\mathbb{E}\{\nu_{\text{y},k}\nu^{\ast}_{\text{x},k'} \}\mathbb{E}\{\xi^{\ast}_{\text{y},m}\xi_{\text{x},n}\xi_{\text{x},m'}\xi^{\ast}_{\text{x},n'} \}+\mathbb{E}\{\nu_{\text{y},k}\nu^{\ast}_{\text{y},k'} \}\right.\\
    &\left.\cdot\mathbb{E}\{\xi^{\ast}_{\text{y},m}\xi_{\text{x},n}\xi_{\text{y},m'}\xi^{\ast}_{\text{x},n'} \}+\mathbb{E}\{\nu_{\text{y},k}\nu^{\ast}_{\text{x},k'} \}\mathbb{E}\{\xi^{\ast}_{\text{y},m}\xi_{\text{x},n}\xi_{\text{y},m'}\xi^{\ast}_{\text{y},n'} \}\right.\\
    &\left.+2\mathbb{E}\{\nu_{\text{x},k}\nu^{\ast}_{\text{x},k'} \}\mathbb{E}\{\xi^{\ast}_{\text{y},m}\xi_{\text{y},n}\xi_{\text{x},m'}\xi^{\ast}_{\text{x},n'} \}+\mathbb{E}\{\nu_{\text{x},k}\nu^{\ast}_{\text{y},k'} \}\right.\\
    &\left.\cdot\mathbb{E}\{\xi^{\ast}_{\text{y},m}\xi_{\text{y},n}\xi_{\text{y},m'}\xi^{\ast}_{\text{x},n'} \}+\mathbb{E}\{\nu_{\text{x},k}\nu^{\ast}_{\text{x},k'} \}\mathbb{E}\{\xi^{\ast}_{\text{y},m}\xi_{\text{y},n}\xi_{\text{y},m'}\xi^{\ast}_{\text{y},n'} \}
    \right].
    \end{split}
\end{equation}

This is now a six-dimensional sum and the complete auto-correlation function consist of nine terms. For ease of analysis, we simply rewrite \eqref{X1} as
\begin{align}
\begin{split}
    & S^{\text{x},h}_{\text{XCI},\text{X}1_i}(f,N_s,L_s)=\left(\frac{8}{9}\right)^2\gamma^2\Delta_f^{3}\sum^{\infty}_{i=-\infty}\delta(f-i\Delta_f)\\ &\cdot\sum_{k,m,n\in \text{X}1_i}\sum_{k',m',n' \in \text{X}1_i}\left\{\mu(\text{X}1_i,N_s,L_s)\mu^{\ast}(\text{X}1_i,N_s,L_s)\right.\\
    &\left.\cdot\sum_{i\in \{0,1,...,W-1\}^6}\left[A_{1,\boldsymbol{i}}(k,m,n,k',m',n')\right.\right.\\
    &\left.\left.+A_{2,\boldsymbol{i}}(k,m,n,k',m',n')
    +A_{3,\boldsymbol{i}}(k,m,n,k',m',n')\right.\right.\\
    &\left.\left.+A_{4,\boldsymbol{i}}(k,m,n,k',m',n')
    +A_{5,\boldsymbol{i}}(k,m,n,k',m',n')\right.\right.\\
    &\left.\left.+A_{6,\boldsymbol{i}}(k,m,n,k',m',n')
    +A_{7,\boldsymbol{i}}(k,m,n,k',m',n')\right.\right.\\
    &\left.\left.+A_{8,\boldsymbol{i}}(k,m,n,k',m',n')
    +A_{9,\boldsymbol{i}}(k,m,n,k',m',n')\right]
    \right\},
\end{split}
\end{align}
where $\boldsymbol{i} \triangleq (i_1,i_2,...,i_6)$ and
\begin{align}\label{A1}
\begin{split}
    A_{1,\boldsymbol{i}}(k,m,n,& k',m',n') \\
    & \triangleq   4\Delta_f^3\left\{\mathbb{E}\{a_{\text{x},i_1}a^{\ast}_{\text{x},i_4} \} \{\mathbb{E}\{b^{\ast}_{\text{x},i_2}b_{\text{x},i_3}b_{\text{x},i_5}b^{\ast}_{\text{x},i_6} \} \right\}\\ &\cdot e^{-j\frac{2\pi}{W}(ki_1-mi_2+ni_3-k'i_4+m'i_5-n'i_6)}.
\end{split}
\end{align}
The other coefficients $A_{j,\boldsymbol{i}}(k,m,n,k',m',n')$ with $j=1,2,...,9$ are similar to $A_{1,\boldsymbol{i}}(k,m,n,k',m',n')$.

Note that under the assumption that the elements in the random vectors $\boldsymbol{a}_n$ and $\boldsymbol{b}_n$ are independent and identically distributed random variables, we have  $\mathbb{E}\{\boldsymbol{a}_n\}=\mathbb{E}\{\boldsymbol{b}_n\}=0$. 
Therefore, some combinations of $(i_1,i_2,i_3,i_4,i_5,i_6)$ result in zero contributions. For example, in the case of $i_1=i_4 \neq i_2=i_3=i_5\neq i_6 $, we have 
\begin{align}\nonumber
\begin{split}
&\mathbb{E}\{a_{\text{x},i_1}a^{\ast}_{\text{x},i_4} \} \{\mathbb{E}\{b^{\ast}_{\text{x},i_2}b_{\text{x},i_3}b_{\text{x},i_5}b^{\ast}_{\text{x},i_6} \}\\=&\mathbb{E}\{|a_{\text{x},i_1}|^2 \}\mathbb{E}\{|b_{\text{x},i_2}|^2b_{\text{x},i_3} \}
\cdot\mathbb{E}\{b_{\text{x},i_6} \}\\=&0.
\end{split}
\end{align}
From this follows that, any combinations of the first-order moment and other-order correlation are zero-contribution combinations. Therefore, we have four possible combinations
\begin{align}\label{combinations}
\begin{split}
  \text{Case}\quad (1) \quad  i_1=i_4 \quad i_2=i_3=i_5=i_6\\
  \text{Case}\quad (2) \quad  i_1=i_4 \quad i_2=i_3 \quad i_5=i_6,\\
  \text{Case}\quad (3) \quad  i_1=i_4 \quad i_2=i_5 \quad i_3=i_6,\\
  \text{Case}\quad (4) \quad  i_1=i_4 \quad i_2=i_6 \quad i_3=i_5.
\end{split}
\end{align}

Here we give a detailed derivation of the term $A_{1,\boldsymbol{i}}(k,m,n,k',m',n')$ in \eqref{A1}. The others terms $A_{j,\boldsymbol{i}}(k,m,n,k',m',n')$ with $j=2,3,...,9$ can be derived following the same approach. As \eqref{combinations} shows, the second-order moment is the set of $i_1=i_4$,
\begin{align}
\begin{split}
    \mathbb{E}\{\nu_{\text{x},k}\nu^{\ast}_{\text{x},k'} \}&=\Delta_f |P(k\Delta_f)|^2\mathbb{E}\{|a_\text{x}|^2\}\sum_{i_1=i_4=0}^{W-1}e^{-j \frac{2\pi}{W}(k-k')i_1}\\
    &=R_s |P(k\Delta_f)|^2\mathbb{E}\{|a_\text{x}|^2\}(\delta_{k-k'-pW}),
\end{split}
\end{align}
where we have used $R_s=W\Delta_f$ and the $p \in \mathbb{Z}$.

Its 4th-order moment is given by
\begin{align}
\begin{split}
    \mathbb{E}\{\xi^{\ast}_{\text{x},m}\xi_{\text{x},n}\xi_{\text{x},m'}\xi^{\ast}_{\text{x},n'} \}=&\Delta_f^2 \mathcal{P}_{mnm'n'}\sum_{i_2,i_3,i_5,_6=0}^{W-1}\mathbb{E}\{b^{\ast}_{i_2}b_{i_3}b_{i_5}b^{\ast}_{i_6} \}\\
    &\cdot e^{-j\frac{2\pi}{W}(-i_2m+i_3n+i_5m'-i_6n')},
\end{split}
\end{align}
where $\mathcal{P}_{mnm'n'}=P_{\text{INT}}^{\ast}(m\Delta+f_\text{c}^h)P_{\text{INT}}(n\Delta+f_\text{c}^h)P_{\text{INT}}(m'\Delta+f_\text{c}^h)P_{\text{INT}}^{\ast}(n'\Delta+f_\text{c}^h)$.

The calculation of the 4th-order moment can be split according to the classification in \eqref{combinations}:
\begin{itemize}
    \item \textbf{Case (1):}
\begin{align}\label{one}
\begin{split}
    \mathbb{E}^{(1)}\{\xi^{\ast}_{\text{x},m}\xi_{\text{x},n}\xi_{\text{x},m'}\xi^{\ast}_{\text{x},n'} \}=
    &R_s\Delta_f \mathcal{P}_{mnm'n'}\mathbb{E}\{|b_\text{x}|^4 \}\\
    &(\delta_{n-m+m'-n'-pW}).
\end{split}
\end{align}
\item  \textbf{Case (2):}
\begin{align}\label{two}
\begin{split}
    &\mathbb{E}^{(2)}\{\xi^{\ast}_{\text{x},m}\xi_{\text{x},n}\xi_{\text{x},m'}\xi^{\ast}_{\text{x},n'} \}=
    \mathcal{P}_{mnm'n'}\mathbb{E}^2\{|b_\text{x}|^2 \}\\
    &(R_s^2\delta_{n-m-pW}\delta_{m'-n'-pW}-R_s\Delta_f\delta_{n-m+m'-n'-pW}).
\end{split}
\end{align}
\item  \textbf{Case (3):}
\begin{align}\label{three}
\begin{split}
    &\mathbb{E}^{(3)}\{\xi^{\ast}_{\text{x},m}\xi_{\text{x},n}\xi_{\text{x},m'}\xi^{\ast}_{\text{x},n'} \}=
    \mathcal{P}_{mnm'n'}\mathbb{E}^2\{|b_\text{x}|^2 \}\\
    &(R_s^2\delta_{m'-m-pW}\delta_{n-n'-pW}-R_s\Delta_f\delta_{n-m+m'-n'-pW}).
\end{split}
\end{align}
\item \textbf{Case (4):}
\begin{align}\label{four}
\begin{split}
    &\mathbb{E}^{(4)}\{\xi^{\ast}_{\text{x},m}\xi_{\text{x},n}\xi_{\text{x},m'}\xi^{\ast}_{\text{x},n'} \}=
    \mathcal{P}_{mnm'n'}|\mathbb{E}\{b_\text{x}^2 \}|^2\\
    &(R_s^2\delta_{-m-n'-pW}\delta_{m'+n-pW}-R_s\Delta_f\delta_{n-m+m'-n'-pW}).
\end{split}
\end{align}
\end{itemize}

Note that we removed the terms with $n=m$ or $n'=m'$ because they have been shown to contribute a frequency-flat and constant phase shift which could be compensated at the receiver. Adding up the contributions in \eqref{one}--\eqref{four}, we obtain the solution of \eqref{A1}.

The $S^{\text{x},h}_{\text{XCI,X1}_i}(f,N_s,L_s)$  is induced by the integration regions X1. As for the other contributions, they can be calculated through the same procedure, and related to different integration regions in Fig.~\ref{fig:WDM}. By combining all the XCI contributions, the XCI PSD can be obtained in x as
\changed{
\begin{align}\label{ss,xci}
\begin{split}
\resizebox{0.95\hsize}{!}{$\begin{aligned}
    S^{\text{x},h}_{\text{XCI}}(f,N_s,L_s)&=\left(\frac{8}{9}\right)^2\gamma^2\Bigg[  \underbrace{R_s^3[\Phi_4\chi^{1}_{\text{XCI,X1}}(f)+\Phi_5\chi^{2}_{\text{XCI,X1}}(f)]
     +R_s^2\Phi_6\chi^{3}_{\text{XCI,X1}}(f)}_{\eta^{\text{x}}_{\text{ss,X1}}}\\
     &+\underbrace{R_s^3[\Psi_5\chi^{1}_{\text{XCI,X2}}(f)+\Psi_6\chi^{2}_{\text{XCI,X2}}(f)]
     +R_s^2\Psi_7\chi^{3}_{\text{XCI,X2}}(f)}_{\eta^{\text{x}}_{\text{ss,X2}}}\\
    &+\underbrace{R_s^3[\Lambda_7\chi^{1}_{\text{XCI,X3}}(f)+\Lambda_8\chi^{2}_{\text{XCI,X3}}(f)]
     +R_s^2\Lambda_9\chi^{3}_{\text{XCI,X3}}(f)}_{\eta^{\text{x}}_{\text{ss,X3}}}\Bigg]
     +\eta^{\text{x},h}_{\text{ss,X4}},
     \end{aligned}$}
\end{split}
\end{align}}
where the $\eta^{\text{x},h}_{\text{ss,X4}}$ is the NLI contribution of X4 region. The X4 region has a similar structure as the SCI region, and thus, we use $\bar{\eta}^{\text{x},h}_{\text{ss,SCI}}$ to denote $\eta^{\text{x},h}_{\text{ss,X4}}$ in \eqref{XCI}. The proof of Theorem~\ref{theorem1} is completed by combining the same kind of terms in \eqref{ss,xci}.

\section{Proof of Theorem~\ref{Theorem2}}\label{sec: appendix_B}
For the NLI contribution of MCI, we take the same approach as in \cite{Carena2014}.  The most important things is to determine the integration regions of the integrand \changed{$\chi^k_{\text{MCI,M1/M2/M3}}, k = 1,2,3$}. In other words, the location of the two three triples $(f_1,f_2,f_3)$ and $(f'_1,f'_2,f'_3)$ need to be determined. Therefore, the MCI PSD can be obtained in x as
\changed{\begin{align}\label{ss,mci}
\begin{split}
\resizebox{0.98\hsize}{!}{$\begin{aligned}
    S^{\text{x}}_{\text{MCI}}(f,N_s,L_s)& = 2\cdot\left(\frac{8}{9}\right)^2\gamma^2{P^3}\Bigg[ \underbrace{R_s^3[\Phi_4\chi^{1}_{\text{MCI,M1}}(f)+\Phi_5\chi^{2}_{\text{MCI,M1}}(f)]+R_s^2\Phi_6\chi^{3}_{\text{MCI,M1}}(f)}_{\eta^{\text{x}}_{\text{ss,M1}}}\\
    &\quad+\underbrace{R_s^3[\Phi_4\chi^{1}_{\text{MCI,M2}}(f)+\Phi_5\chi^{2}_{\text{MCI,M2}}(f)]+R_s^2\Phi_6\chi^{3}_{\text{MCI,M2}}(f)}_{\eta^{\text{x}}_{\text{ss,M2}}}\\
    &\quad+\underbrace{R_s^3[\Lambda_7\chi^{1}_{\text{MCI,M3}}(f)+\Lambda_8\chi^{2}_{\text{MCI,M3}}(f)]+R_s^2\Lambda_9\chi^{3}_{\text{MCI,M3}}(f)}_{\eta^{\text{x}}_{\text{ss,M3}}}\Bigg]+\bar{\eta}^x_{\text{M}0},
\end{aligned}$}
\end{split}
\end{align}}where the main difference between MCI components and their similar XCI components is the
integration limits. Under the assumption that the total number of channels is odd and all INT channels are sitting symmetrically about COI, the INT channels can be expressed as $\text{INT}_h, h = -(N_{ch}-1)/2,...,-1,1,...,(N_{ch}-1)/2$. Due to the symmetry, we will derive a formula for one quadrant, and then multiply it by two. The M1, M2 and M3 are shown as follows:
\begin{itemize}
    \item M1: similar to X1\\
    \changed{We evaluate M1 in the domains locating in the second quadrant, parallel to $f_2$.} We  obtain:\\
    \vspace{-3em}
    \begin{center}
    \begin{align*}
    f_1, f'_1 \in \text{INT}_{-1}, \quad f_2, f_3, f'_2, f'_3 \in \text{INT}_h,
    \end{align*}
    \\ \vspace{-2.5em}
    \begin{align*}
        h = 1, 2,..., (N_{ch}-1)/2
    \end{align*}
    \end{center}    
    \item M2: similar to X1\\
    For the domains locating in the first quadrant, parallel to $f_2$. We obtain:\\
    \vspace{-3em}
    \begin{center}
    \begin{align*}
    f_1, f'_1 \in \text{INT}_1,\quad f_2, f_3, f'_2, f'_3 \in \text{INT}_h,
    \end{align*}
    \\ \vspace{-2.5em}
    \begin{align*}
        h = 2, 3,..., (N_{ch}-1)/2
    \end{align*}
    \end{center}
    \item M3: similar to X3\\
    For the domains locating in the first quadrant, we obtain:\\
    \vspace{-3em}
    \begin{center}
    \begin{align*}
        f_3, f'_3 \in \text{INT}_{h'},\quad f_1, f_2, f'_1, f'_2 \in \text{INT}_{h},
    \end{align*}
    \begin{align*}
    & h' =  2, 3,..., (N_{ch}-1)/2\\
    & h =\left\{
    \begin{aligned}
    & h'/2, h'\text{ is even}\\
    & (h' \pm 1)/2, h' \text{ is odd}
    \end{aligned}
    \right.  
    \end{align*}
    \end{center}
\end{itemize}
The proof of Theorem~\ref{Theorem2} is completed by combining \eqref{ss,mci} with Table~\ref{tab:two} and Table~\ref{tab:three}.

\changed{
\section{Outer boundaries of the integration domain}\label{sec: appendix_C}
The integration regions in Table.~\ref{tab:three} are carried out over the plane $[f_1, f_2]$, which are cubical. But it can be verified that the cubical regions effectively reduce to the lozenge-shaped regions in Fig.~\ref{fig:WDM} due to the support of the integrands. In Fig.~\ref{fig:Outer_boundaries}, we take the SCI as an example to show the outer boundaries. Note that the $f_3$  obeys the fixed relation $f_3 = f_1 + f_2 - f$, where the $f = 0$ for convenience. 
}
\changed{
\begin{figure}[!h]
\vspace{-1em}
    \centering
     \scalebox{0.96}{\includegraphics{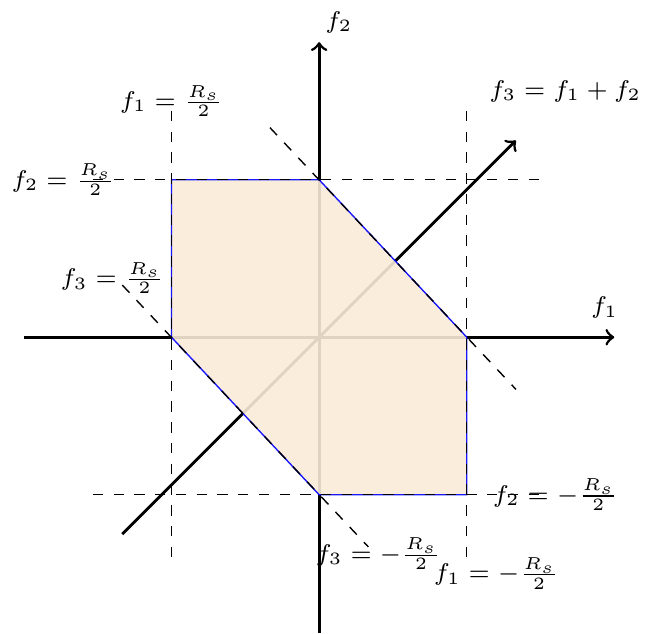}}
    \vspace{-0.4em}
    \caption{\changed{Outer boundaries of the integration of the SCI at the frequency $f = 0$.}}
    \label{fig:Outer_boundaries}
    \vspace{-0.3em}
\end{figure}
}
\end{appendices}

\ch{
\section*{Acknowledgment}
The authors greatly acknowledge the anonymous reviewers who have helped to improve the paper significantly.}

\bibliographystyle{IEEEtran}
\bibliography{references_4D64PRS,references}

\end{document}